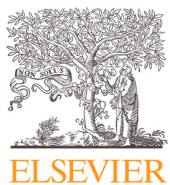

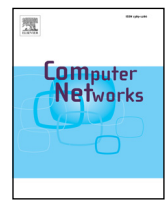

# 5G network slicing using SDN and NFV: A survey of taxonomy, architectures and future challenges

Alcardo Alex Barakabitze [a,*], Arslan Ahmad [b], Rashid Mijumbi [c], Andrew Hines [d]

[a] School of Computing, Electronics and Mathematics, University of Plymouth, UK
[b] IS-Wireless, Poland
[c] Nokia Bell Labs, Dublin, Ireland
[d] School of Computer Science, University College Dublin, Ireland

A B S T R A C T

The increasing consumption of multimedia services and the demand of high-quality services from customers has triggered a fundamental change in how we administer networks in terms of abstraction, separation, and mapping of forwarding, control and management aspects of services. The industry and the academia are embracing 5G as the future network capable to support next generation vertical applications with different service requirements. To realize this vision in 5G network, the physical network has to be sliced into multiple isolated logical networks of varying sizes and structures which are dedicated to different types of services based on their requirements with different characteristics and requirements (e.g., a slice for massive IoT devices, smartphones or autonomous cars, etc.). Softwarization using Software-Defined Networking (SDN) and Network Function Virtualization (NFV)in 5G networks are expected to fill the void of programmable control and management of network resources.

In this paper, we provide a comprehensive review and updated solutions related to 5G network slicing using SDN and NFV. Firstly, we present 5G service quality and business requirements followed by a description of 5G network softwarization and slicing paradigms including essential concepts, history and different use cases. Secondly, we provide a tutorial of 5G network slicing technology enablers including SDN, NFV, MEC, cloud/Fog computing, network hypervisors, virtual machines & containers. Thidly, we comprehensively survey different industrial initiatives and projects that are pushing forward the adoption of SDN and NFV in accelerating 5G network slicing. A comparison of various 5G architectural approaches in terms of practical implementations, technology adoptions and deployment strategies is presented. Moreover, we provide a discussion on various open source orchestrators and proof of concepts representing industrial contribution. The work also investigates the standardization efforts in 5G networks regarding network slicing and softwarization. Additionally, the article presents the management and orchestration of network slices in a single domain followed by a comprehensive survey of management and orchestration approaches in 5G network slicing across multiple domains while supporting multiple tenants. Furthermore, we highlight the future challenges and research directions regarding network softwarization and slicing using SDN and NFV in 5G networks.

## 1. Introduction

The exponential growth of mobile video services (e.g., YouTube and Mobile TV) on smart devices and the advances in the Internet of Things (IoT) have triggered global initiatives towards developing the fifth-generation (5G) mobile and wireless communication systems [1,2,3]. The increasing number of smart devices (e.g., tablets and smartphones) and the growing number of bandwidth-hungry mobile applications (e.g., live video streaming, online video gaming) which demand higher spectral efficiency than that of 4G systems are posing significant challenges in 5G. The Cisco Visual Networking Index (VNI) Forecast [4] predicts that IP video traffic will be 82% of all consumer Internet traffic by 2022, up from 75% in 2017. Mobile video traffic alone will account for 78% of the global mobile data traffic. While the traffic for virtual/augmented reality

* Corresponding author.
*E-mail address:* alcardoalex.barakabitze@plymouth.ac.uk (A.A. Barakabitze).

(VR/AR) will increase at a Compound Annual Growth Rate (CAGR) of 82% between 2017 and 2022, the traffic growth rates of TVs, tablets, smartphones, and M2M modules will be 21%, 29%, 49%, and 49%, respectively. Such a tremendous growth will be the result of 12.3 billion mobile-connected devices, which is expected to even exceed the world's projected population of 8 billion by 2022. A 5G connection is expected to generate 4.7 times more data than that of 4G [4].

With the increasing number of new applications beyond personal communications, mobile devices will probably reach hundreds of billions till the commercial deployment of 5G networks. The 5G network systems around 2020 and beyond will need to deliver as much as 1000 times capacity compared to the current commercial 4G cellular systems [2,5]. The Key Performance Indicators (KPIs) of 5G are expected to include: better, ubiquitous and increased coverage of almost 100% coverage for "anytime anywhere" connectivity, 10–100 times higher user data rates, above 90% energy savings, an aggregate service reliability and availability of 99.999%, an End-to-End (E2E) over-the-air latency of less than 1$ms$ and lowered electro-magnetic field levels compared to LTE [2,6]. The 5G has been triggered by increasing strong demand of a well-connected society context with smart grid and smart cities, critical infrastructure systems such as e-health and telemedicine as well as education sectors which are surging to exploit the total benefits of wireless connectivity by 2020. While 5G is expected to enable the global economic output of $12.3 trillion by 2035 [7], some of the 5G market drivers include the needs for virtual reality, rich media services such as video gaming, 4K/8K/3D video, and applications in smart cities, education and public safety [8]. Industry and academia are embracing 5G as the future network that will enable vertical industries with a diverse set of performance and service requirements. The 5G "theme" has captured attention and imaginations of researchers and engineers around the world with preliminary discussions, debates and a variety of questions such as: (a) What will 5G be? [3] (b) What are the potential technology-enablers and requirements for 5G networks? [2] (c) What are the challenges of 5G? [5], (d) how, and to what extent can future 5G network management be automated to ensure that different service requirements and Experience Level Agreement (ELAs)[1] are fulfilled in the cloud/heterogeneous-native supported softwarized environments [10,11]? (e) How to incorporate the driving system-level principles (e.g., flexibility and programmability) that will allow implementing the vision of 5G network/infrastructure/resource sharing/slicing across network softwarization technologies (SDN, NFV, and MEC)? (f) How to allow and perform dynamic and flexible creation as well as operational control of both Virtual Networks (VNs) and its underlying 5G infrastructure resource pool? (g) What is the disruptive network architecture that can harness all available network technologies and new services to address the 5G challenges?

Although the vision and targets of 5G are clear, the research questions regarding the infrastructure of 5G networks, the enabling technologies, and application scenarios remain open. This attracts global efforts and initiatives from government, organizations, academia and important industry for providing innovative solutions and tackling the critical research questions mentioned above. One of the disruptive concepts that could provide answers to these questions and realize the 5G vision is network slicing (NS) [12,13]. With NS, a single 5G physical network has to be sliced into multiple isolated logical networks of varying sizes and structures dedicated to different types of services. According to the Global

System for Mobile Communications (GSMA) report [14], network slicing is an integral component to unlocking the enterprise opportunity amounting to $300 billion by 2025 for the 5G era. Network slicing will give operators capabilities of creating different level of services for different enterprise verticals, enabling them to customize their operations [14]. However, one of the significant questions is how to meet the requirements of different verticals over 5G networks. This paper provides preliminary answers to some of the above open questions by giving a comprehensive survey of 5G network slicing using SDN and NFV.

### 1.1. Related work and open questions

Following the conception of network slicing, different works in the past have been proposed to identify the potential approaches, uses cases, architectures and the huge benefits brought by network slicing technology in meeting the demands of vertical applications in 5G networks [10,16,17,19,23–29].

Casellas et al. [26] present a control, management, and orchestration of optical systems. Muñoz et al. [25] describe an integrated SDN/NFV-based management and orchestration architecture for dynamic deployment of instances of Virtual Tenant Networks (VTN). Richart et al. [16] provide a review of resource slicing in virtual wireless networks by analyzing SDN and NFV for network slicing. An analysis of 5G network slicing with a focus on the 3GPP standardization process is given in [17]. Habibi et al. [30] provide a discussion on the concept and a system architecture of network slicing with particular focus on its business aspect and profit modeling. The two different dimensions of profit modeling are discussed including (a) Own-Slice Implementation and, (b) Resource Leasing for Outsourced Slices. Foukas et al. [19] presented a survey of network slicing in the 5G context and identify some challenges regarding service-oriented 5G. Yousaf et al. [10] presented the design of a flexible 5G architecture for network slicing with an emphasis on techniques that ultimately provide flexible service-tailored mobility, service-aware Quality of Service (QoS) or Quality of Experience (QoE) control as well as efficient utilization of substrate resources for slicing. A survey of proposals that exploits softwarization and virtualization for the network design and functionality implementation of 5G networks is presented by Massimo et al. [23].

While recent efforts in [24,28,29] provide the description of 5G network slicing in the aspects of SDN/NFV, we note that, these works are limited in at least one of the following: (1) They provide limited review and standardization activities related to 5G network slicing, (2) No comprehensive descriptions of ongoing research projects, State-of-the-Art (SotA) efforts and challenges as well as concrete research directions on how SDN, NFV and Cloud/edge computing can accelerate and exploit the 5G network slicing transformation with embedded intelligent techniques, and (3) With regard to scope, they do not provide important aspects of SDN and NFV for 5G network slicing such as different architectural approaches, their implementations and deployment strategies. Table 1 indicates a summary of related survey papers on network softwarization and 5G network slicing.

### 1.2. Scope and contributions

The major objectives of this paper are to give the reader a comprehensive state-of-the-art and updated solutions related to 5G network slicing using SDN and NFV. We first provide the 5G service quality and business requirements, the description of 5G network softwarization and slicing concepts and different use cases. We also describe standardization activities and different industrial initiatives and projects pushing forward the implementation of 5G network slicing. We summarize our contributions as follows:

---

[1] Experience Level Agreements (ELAs): Indicate a QoE-enabled counter piece to traditional QoS-based Service Level Agreements (SLA) that conveys the performance of the service in terms of QoE. The ELAs establish a common understanding of an end-user experience on the quality levels whiling using the service [9].

**Table 1**
A Summary of Related Survey Papers on Network Softwarization and 5G Network Slicing.

| Contributions and covered scope | [15]-2016 | [16]-2016 | [17]-2016 | [18]-2017 | [19]-2017 | [20]-2018 | [21]-2018 | [22]-2018 | [23]-2018 | Our paper-2019 |
|---|---|---|---|---|---|---|---|---|---|---|
| 5G Service Quality Requirements | ✗ | ✗ | ✗ | ✗ | ✓ | ✓ | ✗ | ✗ | ✗ | ✓ |
| 5G Market Drivers & Key Vertical Segments | ✗ | ✗ | ✓ | ✗ | ✗ | ✓ | ✗ | ✗ | ✗ | ✓ |
| Network Softwarization | ✗ | ✓ | ✓ | ✓ | ✓ | ✓ | ✓ | ✓ | ✓ | ✓ |
| 5G Networks Considerations | ✗ | ✗ | ✓ | ✓ | ✓ | ✓ | ✓ | ✓ | ✓ | ✓ |
| Network Slicing concepts, history and principles | ✗ | ✓ | ✗ | ✗ | ✗ | ✓ | ✗ | ✓ | ✗ | ✓ |
| Virtualization Hypervisors | ✓ | ✗ | ✗ | ✗ | ✗ | ✓ | ✗ | ✗ | ✗ | ✓ |
| Placement of Virtual Resources and VNFs | ✗ | ✓ | ✗ | ✓ | ✓ | ✗ | ✓ | ✓ | ✓ | ✓ |
| 5G Network Slicing Standardization Efforts | ✗ | ✗ | ✗ | ✗ | ✗ | ✗ | ✗ | ✗ | ✗ | ✓ |
| 5G network slicing PoC | ✗ | ✗ | ✗ | ✗ | ✗ | ✗ | ✗ | ✗ | ✗ | ✓ |
| 5G Collaborative Projects | ✗ | ✗ | ✗ | ✗ | ✗ | ✗ | ✗ | ✗ | ✗ | ✓ |
| Orchestrators for Network Slices | ✗ | ✗ | ✗ | ✗ | ✗ | ✗ | ✗ | ✗ | ✗ | ✓ |
| Multi-Domain Orchestration and Management | ✗ | ✗ | ✓ | ✗ | ✓ | ✓ | ✗ | ✗ | ✗ | ✓ |
| Single-Domain Orchestration and Management | ✗ | ✓ | ✗ | ✓ | ✓ | ✗ | ✗ | ✓ | ✓ | ✓ |
| Network Slicing Management in MEC and Fog | ✗ | ✗ | ✗ | ✗ | ✗ | ✗ | ✗ | ✗ | ✗ | ✓ |
| RAN Slicing | ✗ | ✓ | ✓ | ✓ | ✓ | ✓ | ✓ | ✓ | ✓ | ✓ |
| 5G Network Slicing Architectures and Implementations | ✗ | ✗ | ✗ | ✗ | ✗ | ✓ | ✗ | ✗ | ✗ | ✓ |

** Proof of Concepts = PoC.
"✓" indicates that the attributes are provided or applicable in the research work.
"✗" indicates that the attributes are unspecified or non applicable in the research work **

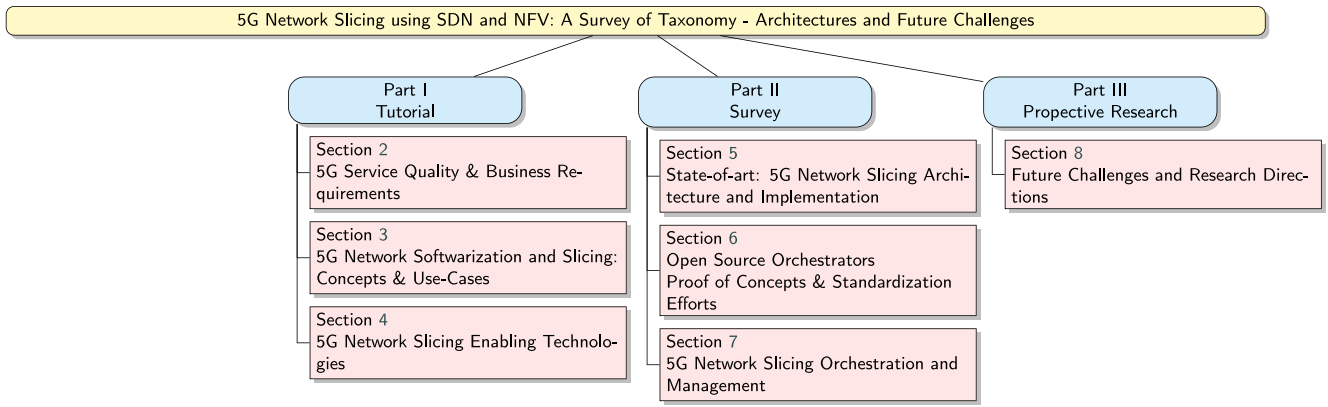

**Fig. 1.** Structure and organization of the paper.

- We describe the prominent service and business requirements for the upcoming 5G network of 2020 and beyond.
- We provide an in-depth discussion on network softwarization along with use cases and scenarios for 5G network slicing.
- We provide a detailed comparison of different SDN/NFV-based architectural approaches and their deployment strategies for 5G network slicing.
- We provide a detailed discussion of standardization activities, research projects and results in network and cloud slicing.
- We further provide a landscape of 5G network slicing orchestration and management in 5G single-domain and multi-domain softwarized infrastructures.
- We further discuss 5G network slicing challenges and explore key research areas in SDN and NFV for future research.

### 1.3. Paper structure and organization

The rest of this paper is organized as follows: we start our discussion with an introduction to the 5G quality of service and business requirements in Section 2. Section 3 introduces the 5G network softwarization and slicing concept, its history and use cases. In Section 4 we present the cutting-edge technologies for enabling the concept of slicing on future 5G networks. In Section 5, we explore different architectures and the state-of-the-art on 5G network slicing from different academic and industry projects. Then in Section 6 we present the open-source orchestrators, proof of concepts (PoC) and standardization activities for 5G network slicing as realized today by the industry and different standard bodies. We provide the convergence and the first realization of SDN and NFV for orchestration and management of 5G network slices in Section 7 in both single-domain and multi-domain environments. We summarize our main findings in Section 8 in the form of future challenges and possible research opportunities before concluding our remarks in Section 9. For a better understanding of the structure and organization of this paper, we refer the reader to Fig. 1. Table 2 provides a list of commonly used acronyms in this paper.

## 2. 5G service quality and business requirements

### 2.1. 5G Service quality requirements

New 5G applications are foreseen to facilitate domains such as M2M, health (e.g., e-health, telemedicine) and education sector. Different 5G applications will need different requirements for their performance. New ways with enhanced capacity (e.g., small cells deployment), intelligent traffic and offload schemes will have to be developed and implemented in order to meet these performance requirements. Moreover, the complexity and high degree of heterogeneity towards 5G also impose the requirements for autonomous network management [31]. Although there are no detailed specifications and general requirements of 5G, exploring 5G requirements (e.g., from users & network perspective) as shown in Fig. 2 that define user's satisfaction in the delivered services is of crucial importance.

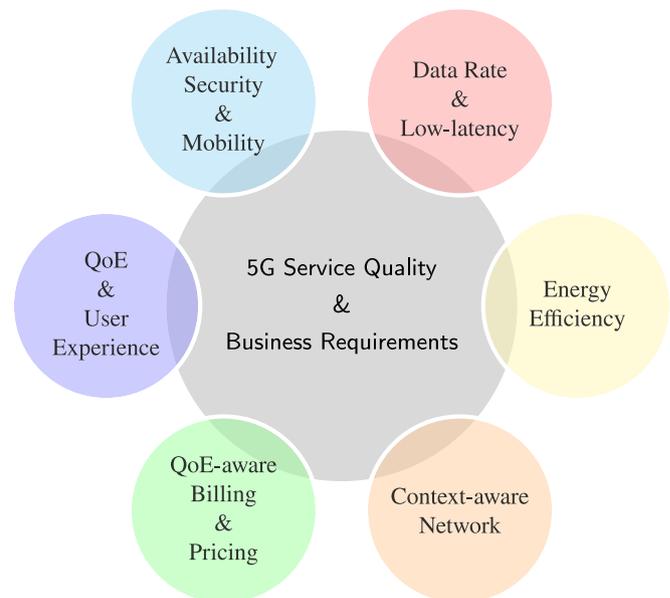

**Fig. 2.** 5G service quality & business requirements [1,3,5].

### 2.2. Data rate and ultra low-latency

The 5G network is expected to provide 1–10 Gbs data rates which are almost ten times of 4G LTE network's theoretical peak data rate of 150 Mbps [1]. With this data rates, 5G will be able to provide a high level of services with guaranteed end-users service quality and a genuinely ubiquitous unlimited mobile broadband experiences even in crowded areas (e.g., stadiums, cars, trains, concerts or shopping malls) through terminals enhanced with Artificial Intelligence (AI) capabilities [32].

5G networks are also envisaged to provide almost 100% coverage for "anytime anywhere" connectivity and a 1ms round trip latency for tactile Internet [33]. In particular, peak data rates in the order of 10 Gb/s will be required to support services such as 3D gaming and mobile telepresence with 3D rendering capabilities [3]. The 5G networks will need to support a higher data rate and deliver higher resolution videos with better QoE to consumers. The reduced latency and high data rate in 5G will easily support high-definition streaming from cloud-based technologies and enhanced VR devices such as Google Glass and other wearable computing devices. It will also provide faster web downloads and enable premium user experience when delivering services, for example, YouTube videos with high-resolution regardless of access method.

**Table 2**
A list of commonly used acronyms in this paper.

| Abb. | Definition | Abb. | Definition | Abb. | Definition |
|---|---|---|---|---|---|
| 5G | Fifth Generation | LSDC | Lightweight Slice Defined Cloud | RLC | Radio Link Control |
| ACTN | Abstraction and Control of Traffic Engineered Networks | M2M | Machine to Machine | RRM | Radio Resource Management |
| B2B | Business-to-Business | MANO | Management and Orchestration | SaaS | Software as a Service |
| B2C | Business-to-Customer | MdO | Multi-domain Orchestrator | SDMC | Software Defined Mobile Network Control |
| BSS | Business Support System | MDSO | Multi-Domain Slice Orchestrator | SDMO | Software-Defined Mobile network Orchestration |
| BSSO | Business Service Slice Orchestrator | MEC | Multi-Access Edge Computing | SFC | Service Function Chaining |
| CAPEX | Capital Expenditure | MIoTs | Massive Internet of Things | SGW | Service Gateway |
| CC | Cloud Computing | MO | Management and Orchestration | SLAs | Service Level Agreements |
| CDNs | CDNs Content Distribution Networks | MTC | Machine Type Communications | SRO | Slice Resource Orchestrator |
| C-RAN | Cloud RAN | MTCP | Mobile Transport and Computing Platform | SBS | Service Broker Stratum |
| D2D | Device to Device | NAT | Network Address Translation | SDO | Standard Developing Organisations |
| DHCP | Dynamic Host Configuration Protocol | NFs | Network Functions | TN | Transport Networks |
| DSSO | Domain -Specific slice Orchestration | NFV | Network Function Virtualization | TOSCA | Topology and Orchestration Specification for Cloud Applications |
| EC2 | Elastic Compute Cloud | NFVIPoP | Point of Presence USDL | NFVI | Point of Presence USDL Universal Service Definition Language |
| ELA | Experience Level Agreement | NFVO | Network Functions Virtualisation Orchestrator | VMN | Virtual Mobile Networks |
| ETSI | European Telecommunication Standard Institute | NGN | Next Generation Networks | VMS | Virtual Machines |
| FoC | Fog Computing | ONF | Open Network Foundation | VNF-FGs | VNF Forwarding Graphs |
| IRTF | Internet Research Task Force | OPEX | operational expenditure | VNFs | Virtual Network Functions |
| ISPs | Internet Service Providers | OSS | Operations Support Systems | VPN | Virtual Private Networks |
| ITU | International Telecommunication Union | PGW | Packet Data Network Gateway | VR/AR | Virtual/Augmented Reality |
| KPR | Key Performance Requirements | PoP | Point of Presence | WWRF | Wireless World Research Forum |
| KQIs | Key Quality Indicators | QoBiz | Quality of Business | XCI | Xhaul Control Infrastructure |
| LAN | Local Area Network | RAN | Radio Access Network | ZOOM | Zero-time Orchestration, Operations and Management |

### 2.3. Enhanced service availability, security and mobility

The 5G needs to be robust enough reliable and resilient network to support timely communications for emergency and public safety. M2M/D2D communicating devices such as smart grid terminals, cars, health monitoring devices, and household appliances will be dominant in 5G network. These devices will need an enhanced service availability with a high-speed connection to the Internet. While today's mobility management protocols are highly centralized and hierarchical [34], 5G network has to cope significantly with such extreme situations by providing mobility on demand based on each device and service's requirements. However, for the full mobility support, enhancements to the current mobility management procedures are needed. For example the handover procedures and a topology-aware gateway selection and relocation algorithm [35]. The new introduced distributed mobility management (DMM) [34,36] proposals for 5G seems to be a solution to overcome the current mobility management limitations.

In terms of security, the current 4G network has limited protection needs on users (e.g., data encryption) and network (e.g., strong authentication for billing). This is different in 5G network which needs to support new business and trust models, new service delivery models with increased privacy concerns and an evolved threat landscape. The 5G network will therefore need to ensure and have the ability towards defending against security attacks such as Denial of Service (DoS) for critical mission applications such as smart grids, public safety, water distribution and natural gas networks [8,37].

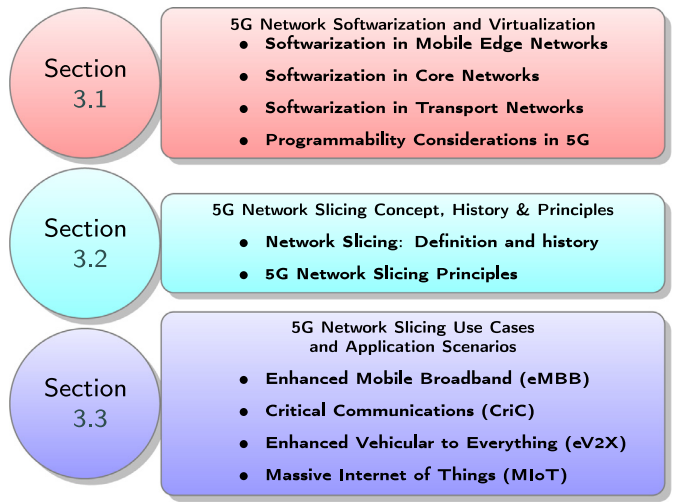

**Fig. 3.** Summary and tutorial contributions of Section 3.

### 2.4. Consistency, transparency, user's QoE personalization and service differentiation

Consistency should be the central requirement for ensuring high level of QoE [38,39] while delivering service to end-users in 5G ecosystem. For 5G to guarantee the required end-user's QoE, the fluctuations in network quality and performance, disruptions and unpredictable interference should be at minimal level. 5G networks should allow high level of transparency in the efforts of delivering services with high QoE to end-users by hiding its complexity. The 5G transparent network has to facilitate the *"Best experience"* for providing an efficient delivery of remote services and data to end-users particularly through cloud data centres hosted by cloud provider's infrastructure.

Different types of 5G applications will need different QoE requirements. For example, various media types have different set of KPI metrics. In this case, service quality differentiation and application type using a personalized QoE management solution are expected in 5G ecosystem [40]. Each user's QoE [38,41,42] and service on the 5G network should be well and autonomously managed. For such QoE personalization, specific modeling of charging mechanisms regarding quality levels, purchases, and content for a user/service have to be developed in 5G networks. Personalizing User Interfaces(UI) in the context of Video-On-Demand/Live-TV services to learn from a user's content consumption patterns can be another approach for QoE personalization on 5G networks [43]. Fortunately, with the development of cloud computing, real-time computations and large-scale online modeling are available, such as Netflix movie and Google advertising [40,44].

### 2.5. Longer battery life, seamless user experience and context aware networking

A billion of cellular-enabled IoT applications involving a battery operated sensor networks will be dominant in 2020 and beyond. The 5G deployment based sensor networks will only be possible if their daily operations will guarantee much longer battery life and the reduced energy consumption of 5G devices for several years [8,45]. With the emerging spectrum bands and the inter-networking among technologies, future 5G networks should be able to deliver and provide a consistent user experience irrespective of the user's location while the quality of achievable latency and data rate being the KPIs. Moreover, 5G solutions should have attributes that will enable the network to adapt to the requirements of connected smart devices and applications. 5G should be a network of varying capabilities with an alternative small cell, multi-RAT and macro networks, with applications and devices QoE requirements.

### 2.6. QoE-based service billing and pricing

QoE-based service billing and pricing are the requirements that have a strong correlation with the end user's perceived quality on 5G systems. The Quality of Business (QoBiz) [46] aspects should be based on well defined QoE-based service billing/charging policies or rules by service providers. For example, a premium IPTV customer who pays more for a service expects a better service quality [47]. Therefore, providing a QoE differentiation in future 5G networks should be concurrently implemented with an appropriate QoE-based service billing and pricing mechanisms that will translate directly into the quality of business.

### 2.7. QoE-rich resource and energy efficient

The base stations (BSs) in 4G LTE networks are inefficient because of their operational cost and high energy consumption. They contribute between 60% and 80% of the whole cellular network energy consumption [48]. While that is the case, mobile video is one of the definitive energy intensive consuming services from user's side. The mobile terminals, for example, consume to around 10% of the total energy consumed by the BSs [48,49]. The reason is that, a mobile video service needs to cooperate with screen display, video/audio decoder, CPU, and network interface. In areas with a high density of users and time variant traffic patterns, 5G should provide an efficient way to optimize the number of active network elements as the traffic grows/decreases and make the network more efficient in terms of energy consumption. The deployment of low-cost low-power access nodes such as small cells [50] and relays has been proposed to be an approach to reduce energy in 5G networks. Such an approach would enable the dynamic resource allocation management that will avoid wastage of energy by adopting different network load variations to crucial network performance indicators/parameters while satisfying the end user's demands. High power consumption of traditional macro BS have triggered researchers and standardization bodies towards designing an energy efficient 5G wireless networks. Projects such as Energy Aware Radio and neTwork tecHnology (EARTH) [51] GreenTouch [52] and Green 5G Mobile Networks (5GrEEn) [53] have already realized and promoted the value of energy-efficient 5G networks.

### 2.8. 5G market drivers & key vertical segments

The industry foresees 5G as the network where different applications and services will be served by a highly integrated and configurable network automatically. In the 5G era, users will merely request services they need, and the information will be delivered to their desired location and device [8,54] without interruptions on service quality. The 5G network is about enabling new services and devices, connecting new industries and empowering new user experiences. This will entail connecting people and things across a diverse set of vertical segments including (a) IoTs for smart grid and critical infrastructure monitoring; (b) smart cities for use cases like smart transportation, smart homes and smart building; (c) M-health and telemedicine; (d) automotive industry for use cases like vehicular Internet/infotainment, cooperative vehicles, inter-vehicle information exchange; (e) media and entertainment (e.g., immersive and interactive media, cooperative production, collaborative gaming). The business roles that are to be facilitated by the upcoming 5G architecture through network softwarization and slicing are summarized in Table 3.

**Table 3**
A summary of 5G network slicing business roles and emerging markets .

| 5G Business Drivers | Business Roles/Objectives |
|---|---|
| Application providers | Offer different applications and services to the end users based on their demands and quality requirements. |
| Vertical markets | Provide different services to third parties that exploit resources (network and cloud) specifically from operators and cloud service providers. |
| Service broker | To map requests coming from application providers, VNO and different industry verticals to MNO's resources. |
| Virtual Network Operators (VNO) | Work with infrastructure providers to offer their telecom services by acquiring the required network capacity to customers. |
| Cloud Providers | Provide computation and storage resources to third parties including cloud resources such as Amazon web service's Elastic Compute Cloud (EC2). |
| Infrastructure Providers | To provide both physical (hardware) and software resources including 5G network connectivity. |

### 2.9. Summary and lesson learned

This section surveys service quality and business requirements in 5G networks. Requirements related to service quality include (a) high data rate (e.g., 1–10 Gbps connections to end points) and 1 ms E2E round trip latency, (b) enhanced service availability (e.g., 99.999% availability and 100% coverage), security and mobility, (c) consistency, transparency, user's QoE personalization and service differentiation, (d) seamless user experience and context aware networking, (e) QoE-based service billing and pricing, and (f) QoE-rich resource and energy efficient (e.g., up to ten years battery life for low power and machine-type devices). The 5G network business drivers and vertical segments are summarized in Table 3. It is important to mention that, new 5G applications and services in 5G networks are foreseen to facilitate domains such as M2M, IoTs for smart grid and smart cities, immersive/interactive media and many more. The new vertical applications and services will need different requirements for their performance. Therefore, new solutions with enhanced capacity (e.g., small cells deployment), intelligent control and management schemes using network softwarization paradigms will have to be developed in order to meet these 5G performance requirements.

## 3. 5G network softwarization and slicing: Concepts & use cases

### 3.1. 5G network softwarization

Network softwarization is an approach that involves the use of software programming to design, implement, deploy, manage and maintain network equipment/components/services [20,55]. Network softwarization aims to deliver 5G services and applications with greater agility and cost-effectiveness. Along with the realization of 5G network requirements (e.g., programmability, flexibility, and adaptability), network softwarization is set to provide E2E service management and improve the end user's QoE [39,56]. Network slicing-as-a-service [57] and the overall 5G E2E service platform unification will be realized by network softwarization, and virtualization using SDN, NFV and cloud computing technologies. The collective expressive power of softwarization and virtualization technologies are the main drivers of innovations in the 5G era where developers and operators can quickly build application-aware networks and network-aware applications to match their business demands. In order to achieve network softwarization goals, new design and implementation is needed in different 5G network seg-ments (e.g., RAN, transport networks, core networks, mobile-edge networks, and network clouds). This is so because each segment has different requirements or technical characteristics and level of softwarization [58]. Fig. 4 illustrates the software network technologies applied in 5G network segments. In the following subsections, we provide an overview of softwarization focusing on RAN, mobile edge networks, core networks, and transport networks.

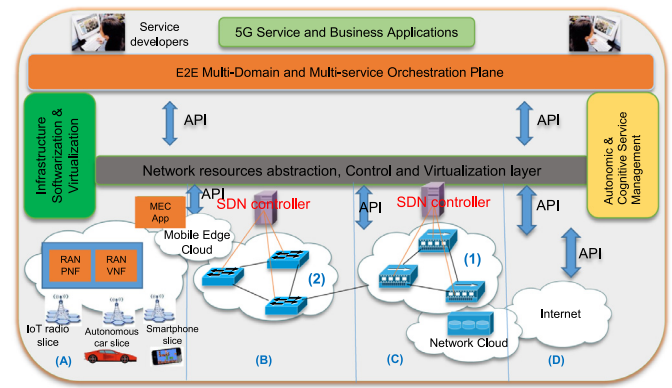

**Fig. 4.** Software network technologies in 5G architecture. A indicates RAN; B = transport networks; C = core networks and D represents the Internet.

### 3.1.1. Softwarization in mobile edge networks

Mobile edge network aims to move contents, network functions, and resources closer to the end-user by extending the conventional data-center to the edge of 5G networks. Softwarization in mobile edge networks will be implemented based on the virtualized platform that leverages SDN, NFV and Information-Centric Networking (ICN) [59,60]. MEC [61] is a new technology with the main idea of implementing a content-oriented and embedded intelligence at the edge in 5G network. Characterized by high bandwidth, low latency, location awareness, and real-time insight radio network information, MEC provides cloud computing capabilities to satisfy high-demanding requirements of 5G such as throughput and an improved QoE for the end-users [62]. Through caching contents at the MEC server, a similar concept to ICN, softwarization of MEC in 5G promises to reduce the volume of data transmitted at the 5G core network for processing, enable real-time and application flow information as well as efficient use of available resources [63].

### 3.1.2. Softwarization in core networks

The design of most core networks and service plane functions in the era of the 5G network are expected to be implemented as VNFs following the envisaged SDN/NFV architectural principles. This will make them run in Virutal Machines (VMs) potentially over standard servers enabled on Fog/Cloud Computing (CC) en-vironments [64,65]. These softwarization capabilities can be deployed at different network sites based on specific service requirements. For example, network slices can use CN and service VNFs based on the required storage capacity and latency of the requested service.

### 3.1.3. Softwarization in transport networks

To adapt to the needs of 5G RANs, future programmable transport networks should be implemented as a platform where various user and network services can be accommodated. The design of such softwarized transport network can be done using appropriate interfaces in SDN/NFV infrastructures. That way, resource discovery, and optimization mechanisms can be easily implemented in the 5G control plane [66]. It is important to mention that, a softwarized 5G transport network will allow for tightly coupled interactions with the RAN where aspects such as mobility and load balancing can be coordinated efficiently [66].

### 3.1.4. Programmability considerations in 5G

Network programmability is a concept that involves network softwarization and virtualization using SDN/NFV infrastructure. 5G programmability needs a systematic splitting and abstraction of NFs to cope with the emerging needs of 5G network efficiency and reliability, service flexibility and security [66]. 5G programmability empowers the fast, flexible and dynamic deployment of new network and management services that can be executed as groups of VMs in all segments of the network (control and management plane). 5G programmability will facilitate the creation of 5G ecosystems that could benefit different control and management planes intuitivelynetwork-wide by utilizing open Application Programming Interface (API) and Software Development Kit (SDK).

## 3.2. 5G network slicing concept, history & principles

### 3.2.1. Network slicing: definition and history

Since 1960s [67] the concept of network slicing has relied heavily on virtualization concepts [68] following the first IBM's operating system (CP-40) design that supported time-sharing and virtual memory. Such a design introduced a system that was able to accommodate up to fifteen users simultaneously [69] and an individual could be allowed to work independently on a separate set of both hardware and software [67,69]. Since then, the idea of network virtualization, where a virtual entity could be created from a physical entity was formed. The vision was to span virtual systems across different network resources, computing infrastructures, and storage devices [68]. In the 1970s and early 1980s, network virtualization was widely adopted in data centers where remote sites were connected with a secured and controlled performance through the Internet.

In the late 1980s, overlay networks were proposed where network nodes were connected over logical links to form a virtual network running over a common physical infrastructure. Overlay networks are an early form of the network slicing concept since it combines different resources over various administrative domains while guaranteeing the QoS to the end-users. Although, overlay networks are flexible, they lack automation and programmability features in the network controls. Throughout the 1990s and in early 2000s, an active and programmable network where a node operating system can provide resource control frameworks was proposed. Since then, different platforms and Federated Testbed (e.g., Planet Lab USA (2002) [70], PlanetLab EU (2005)[2], OneLab EU (2007)[3], PlanetLab Japan (2005), OpenLab EU (2012)[4]) where new network protocols can be verified and evaluated were established. For example, PlanetLab [70,71] adopted a common software package called MyPLC[5] that enables a distributed virtualization where users can obtain slices for specific applications. In 2008, a US National Science Foundation (NSF)[6] project, introduced a GENI [72] testbed based on network virtualization concepts. The aim was to promote research on a clean slate network while considering federated resources and mobile network environments. GENI[7] is a shared network testbed where multiple experimenters may be running multiple experiments at the same time. Following this trend, in 2009, SDN [73] enabled researchers to run experiments in a network slice of a campus network where capabilities of programmability were employed through open interfaces [74].

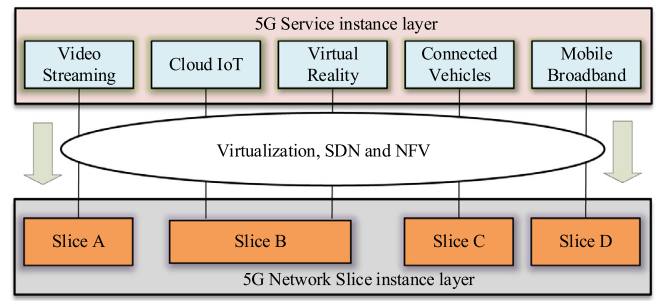

**Fig. 5.** The NGMN network slicing concept.

### 3.2.2. 5G network slicing principles

5G network slicing was coined and first introduced by the Next Generation Mobile Network (NGMN) [75]. As defined by the NGMN, a network slice is an E2E logical network/cloud running on a common underlying (physical or virtual) infrastructure, mutually isolated, with independent control and management that can be created on demand. A network slice may consist of cross-domain components from separate domains in the same or different administrations, or components applicable to the access network, transport network, core network, and edge networks. Network slices are therefore self-contained, mutually isolated, manageable and programmableto support multi-service and multi-tenancy. Fig. 5 represents the NGMN slice capabilties which consists of following three layers [45]:NGMN slice capabilities [45] as shown in Fig. 5 consists of 3 layers as described below:

- 5G Service Instance Layer (5GSIL): represents different services which are to be supported. A Service Instance represents each service. Typically, services can be provided by the network operator or by third parties.
- 5G Network Slice Instance (5GNSI): provides network characteristics which are required by a 5GSI. A 5GNSI may also be shared across multiple 5GSIs provided by the network operator. The 5GNSI may be composed by none, one or more sub-network instances, which may be shared by another NSI.
- 5G Resource Layer (5GRL): It consists of physical resources (asset for computation, storage or transport including radio access) and logical resources (partition of a physical resource or grouping of multiple physical resources dedicated to a Network Function (NF)[8] or shared between a set of NFs).

---

Network slicing concept can facilitate multiple logical and self-contained networks on top of a shared physical infrastructure platform [76]. Since then, different standardization bodies have explored the definition of network slicing from a different perspective.

The ITU envisage network slicing as the basic concept of network softwarization that facilitates a Logical Isolated Network Partitions (LINP) composed of multiple virtual resources, isolated and equipped with a programmable control and data plane [77]. The 3GPP [78] defines network slicing as a *"technology that enables the operator to create networks, customized to provide optimized solutions for different market scenarios which demand diverse requirements (e.g., in terms of functionality, performance, and isolation)"* [79]. From a business perspective, a slice includes a combination of all relevant network resources, functions, and assets required to fulfill a specific business case or service, including OSS, BSS and DevOps processes. As such there are two types of slices: (a) internal slices, understood as the partitions used for internal services of the provider, retaining full control and management of them, (b) external slices, being those partitions hosting customer services, appearing to the customer as dedicated networks/clouds/data-centers.

Network slicing can offer radio, cloud and networking resources to application providers or different vertical segments who have no physical network infrastructure. That way, it enables service differentiation by customizing the network operation to meet the requirements of customers based on the type of service [80]. Basic principles that encompass network slicing and its related operation on 5G softwarized networks are the following: [24,76,81]:

- *Automation of network operation*: Automation allows dynamic life-cycle management of network slices (e.g., deploying, changing, deleting), optimization of network resources (auto-scaling/migration/auto-healing) as well as a dynamic interplay between management and data planes [76].
- *High-Reliability, Scalability and Isolation*: These are the major features of 5G network slicing that ensures performance guarantees and security for each tenant using immediate fault detection mechanisms for services with different performance requirements [21].
- *Programmability*: Programmability simplifies the provisioning of services, manageability of networks and integration and operational challenges especially for supporting communication services [83]. For example, it allows third parties to control the allocated slice resources (e.g., networking and cloud resources using open APIs that expose network capabilities. This, in turn, facilitates on-demand service-oriented customization and resource elasticity on 5G softwarized and virtualized networks [84].
- *Hierarchical Abstraction*: Network slicing introduces an additional layer of abstraction by creating logically or physically separate groups of network resources and (virtual) NFs configurations [75]. This abstraction facilitates service provision from a network slice service on top of the prior one. For example, network operators and ISP can exploit network slicing to enable other industrial companies to use networks as a part of their services (e.g., vertical players like a connected car with the highly reliable network, an online game with ultra-low latency, video streaming with guaranteed bandwidth, etc.) [14].
- *Slice customization:* Slice customization is realized at all layers of the abstracted network topology using SDN that decouples the data and control plane. On the data plane, NFV capabilities described in Section 4.3 provides service-tailored NFs and data forwarding mechanisms where value-added services can be enabled using Artificial Intelligence (AI). It is worth mentioning that, customization assures network resources allocated to a particular 5G tenant are efficiently utilized in order to meet the requirements of a particular service [85].
- *Network Resources Elasticity*: The elasticity of network resources is realized through an effective and non-disruptive re-provisioning mechanism where the allocated resources are scaled up/down. As such, elasticity ensures that the desired SLA/ELAs of users regardless of their geographical location is achieved [86].

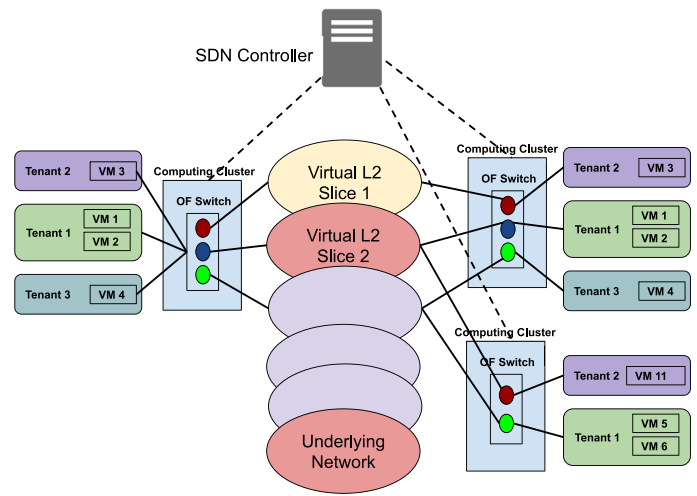

**Fig. 6.** Network slicing use case [82].

### 3.3. 5G network slicing use cases and application scenarios

Aligned with the anticipated NGMN industrial vision for 5G as summarized in [75] to address several emerging services and business demands beyond 2020, the 3GPP initiated a study named New Services and Market Enablers (SMARTER) [91] in the 3GPPP Services Working Group SA 1. More than 70 use cases focusing on new market segments and different business opportunities that could be launched with the arrival of 5G were identified and grouped into the following main categories summarized in Table 4.

As an example, Fig. 6 shows an application scenario of network slicing. A tenant in this context is defined as a logical entity that owns and operate either one or more Virtual Infrastructures (VIs) or network services. That way, it can allocate VIs over its substrate network and provide multiple L2 network slices to offer services to different tenants. Each tenant such as a Mobile Virtual Network Operator (MVNO) owns and operates a network slice. In that aspect, virtual L2 slice 1 is owned by tenant 1 and tenant 2. It is important to note that, other tenants can share any tenant's infrastructure.

The MVNO tenants can, therefore, deploy their network services or allow multiple third-party tenants, for example, over-the-top (OTT) or service providers to instantiate their services on top of the VI. Following a recursive approach [92], it is possible to instantiate a VI on top of another one. That way, the VI of tenant 2 can be instantiated over the VI of tenant 1. As shown in Fig. 6, the SDN controller maintains and coordinates tenant's access to the shared infrastructure and drive resource allocation for instances that are assigned to different tenantswhich can enable the delivery of multi-tenancy related services using dedicated APIs. Similar to the ETSI NFV MANO proposal, the controller manages the logical mapping between tenants, assigned services (in terms of VNFs instances) and the underlying virtual resources allocations, in compliance with the established ELAs/SLAs [26,85,93].

**Table 4**
A summary of 5G Network Slicing Use Cases and Application Scenarios .

| 5G Use Case | Contribution/Objectives/Functionality |
|---|---|
| Enhanced Mobile Broadband (eMBB) | To provide high data rates on 5G systems so as to cope with huge data traffic volumes and UE connectivity per area [87]. |
| Critical Communications (CriC) | To facilitate mission critical services such as the tactile Internet [33], public safety, disaster and emergency response [88] and AR/VR. |
| Enhanced Vehicular to Everything (eV2X) | Focuses on safety-related services such as remote driving, vehicle platooning, autonomous and cooperative collision avoidance by allowing direct vehicular communications [89]. |
| Massive Internet of Things (MIoT) | To provide a common communication conenctivity and inter-networking for various smart devices in the area of smart cities, smart homes and smart farming [90]. |

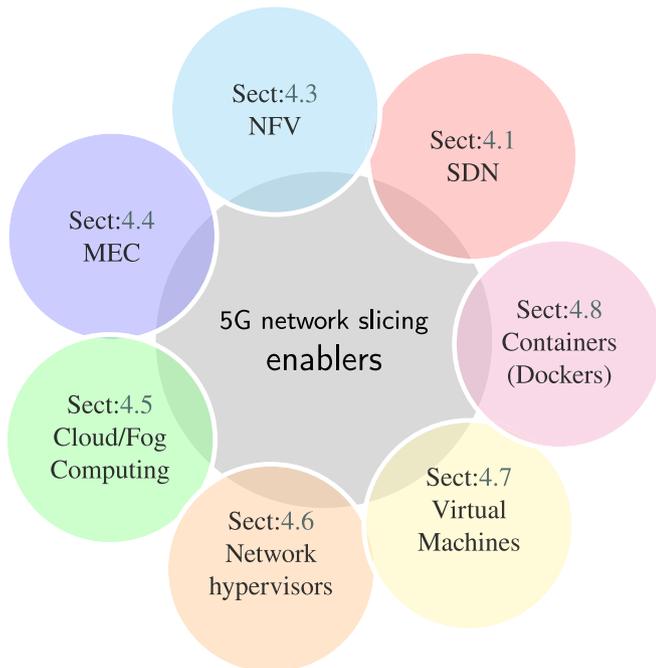

Fig. 7. A summary and a tutorial contribution of Section 4.

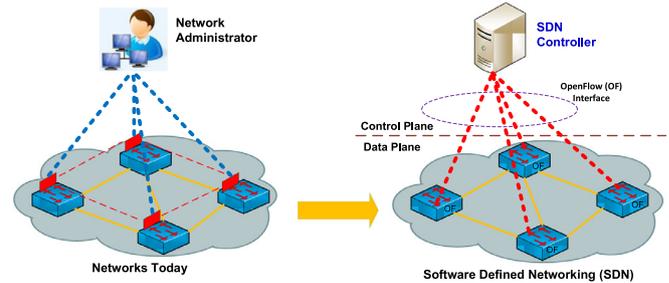

Fig. 8. A comparison of SDN and network operation today.

### 3.4. Summary and lesson learned

This section presents the concepts of network softwarization and slicing including their history and operational principles as summarized in Fig. 3. We include the softwarization mechanisms in mobile edge networks, core and transport networks using promising technologies such as SDN, NFV and MEC. The basic principles of 5G network slicing include automation of network operation, high-reliability, scalability, isolation, and programmability, hierarchical abstraction and slice customization as well as network resources elasticity. We summarize main groups of 5G network slicing and use case scenarios such as that shown in Fig. 6. We note that, 5G network softwarization and slicing is set to facilitate future network management and orchestration of resources from service providers to the end-users. The E2E multi-domain and multi-tenancy support in 5G network slicing promise to provide services across multiple network segments and different administrative domains such that one slice can combine resources belonging to distinct infrastructure providers. Multi-domain aspects in 5G network slicing will also enable to unify different network layers and different technologies from RAN, core network, cloud transport networks [94]. Moreover, the network slicing may enable service oriented network automation via 5G network technology enablers such as SDN, NFV and MEC. Additionally, network slicing may cre-ate new market opportunities for the network providers in future such as offering "Network as a Service" to third party. However, it may also create research challenges regarding new techniques and algorithms for network resource management in the virtualized networks.

## 4. 5G network slicing enabling technologies

### 4.1. Software defined networking (SDN)

SDN is an approach that brings intelligence and flexible programmable 5G networks capable of orchestrating and controlling applications/services in more fine-grained and network-wide manner [73,95]. The Open Network Foundation (ONF) [96] defines SDN as "*the physical separation of the network control plane from the forwarding plane, and where a control plane controls several devices*". This separation results into flexibility and centralized control with a global view of the entire network. It also provides capabilities of responding rapidly to changing network conditions, business, market and end user needs. As shown in Fig. 8, SDN creates a virtualized control plane that can enforce intelligent management decisions among network functions bridging the gap between services provisioning and network management. With SDN, the network control becomes directly programmable using standardized Southbound Interfaces (SBI) such as OpFlex [97], FoRCES [98] and OpenFlow [74]. These standards define the communication between forwarding devices in the data plane and the elements in the control and management plane. The forwarding plane of SDN can be implemented on a specialized commodity server [99] such as VMware's NSX platform [100] which consists of a controller and a virtual switch (vSwitch).

However, such implementations depend on the performance needs and capacity requirements of SDN environments. Strictly narrating, the academia, industry and standard bodies such as the ONF, the Software Defined Networking Research Group (SDNRG) of the Internet Research Task Force (IRTF) and the Internet Engineering Task Force (IETF) have already realized the potential of SDN and defined its architectural components, interfaces and functional requirements for the future 5G networks [39]. SDN is set to ad-dress limitations of the traditional networks (see Fig. 8) which are ill-suited for the dynamic network configuration, control, and management as well as storage needs for today's data centers, campuses, and heterogeneous environments. The SDN paradigm for 5G network slicing analysis is elaborated comprehensively by the ONF [86]. Every SDN client context in the ONF architecture indicates a potential slice as shown in Fig. 9. The SDN controller manages network slices using a set of rules or policies. The SDN controller facilitates the creation of both server and client contexts as well as the installation of their associated policies [12]. In particular, the SDN controller maintains a network slice client context. That way, it allows an SDN controller to dynamically manage network slices by grouping slices that belong to the same context [101]. The SDN controller governs its slices and performs resource orchestration on the server context. The client context consists of support, client and virtual resources to satisfy any incoming requests from end users. Table 5 shows some of SDN solutions that can support network slicing in 5G systems.

**Table 5**
A summary of SDN controllers for network orchestration and dynamic network slicing.

| SDN Controller | Contributions/Objectives/Functionality |
| --- | --- |
| Mobile Central Office Re-architected as a Datacenter (M-CORD) [102] | M-CORD is a cloud-native solution that employs SDN, NFV to provide services to carriers deploying 5G mobile wireless networks. The RAN programmability and virtualization acts as a building blocks for E2E slicing in M-CORD. |
| The Open Network Operating System (ONOS) [103] | ONOS can enable the network slicing concept through VNF composition in the central office where tenants can easily create network services using northbound abstractions. |
| OpenDayLight (ODL) | ODL is set to provide dynamic services in the era of 5G by optimizing softwarized and virtualized networks in order to meet the continuously evolving service demands from the end-users. |

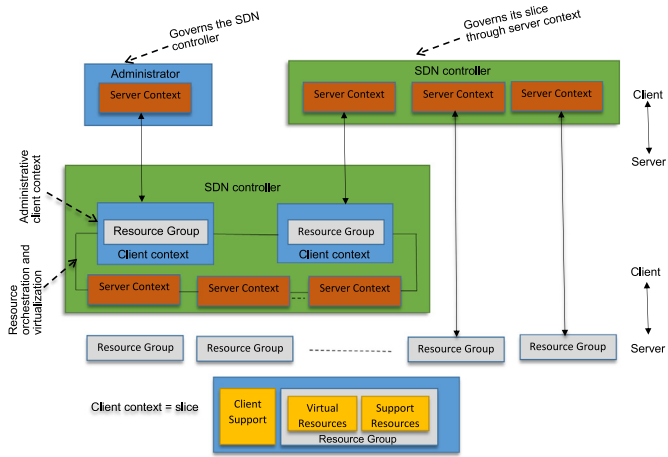

**Fig. 9.** ONF SDN network slicing architecture [86].

### 4.2. Traffic management applications for stateful SDN data plane

Traffic management in SDNs is achieved by OpenFlow which provides a platform-agnostic programmatic interface between the data plane and control plane. OpenFlow focuses solely on L2/L3 network transport and it dynamically updates the match/action forwarding rules only via the explicit involvement of an external controller. Although the OpenFlow specification contains multiple flow tables in the OpenFlow pipeline, it cannot maintain state information in the SDN data plane. OpenFlow also relies heavily on the SDN controller to maintain the states of all packets [104,105]. Such static nature of the OpenFlow forwarding abstraction could raise scalability, reliability and security problems in 5G network slicing because of the control channel bottleneck and pro-cessing delay imposed between the SDN controller and switches [105]. Thanks to the advanced switch interface technologies such as OpenState [106], P4 [107], POF [108], Stateful Data Plane Architecture (SDPA) [109] and SNAP [110] that provide enhanced stateful forwarding and expose persistent state on the SDN data plane [106]. P4 is a high-level language for programming protocol independent that enables programmers to change the way SDN switches process packets.

The advanced data plane programmability (ADPP) enhances the network softwarization capabilities with more agility and flexibility to meet the requirements of 5G network slicing. The ADPP would allow developers to fully exploit resources of SDN data plane for their 5G network applications [111]. Furthermore, it will support resource slicing and isolation as well as facilitating an efficient and automated deployment of new 5G network services over the programmable SDN data plane. With stateful forwarding technologies, the network slices of softwarized 5G architecture are required to be monitored, controlled and managed independently while supporting diversified protocols and data transport mechanisms [111].

### 4.3. Network function virtualization (NFV)

NFV [84] is the virtualization of network functions (e.g., Firewalls, TCP optimizers, NAT64, VPN, DPI) on top of commodity hardware devices. NFV envisages the instantiation of VNFs on commodity hardware. This way, it breaks the unified approach to use software and hardware that exists in traditional vendor offerings. With NFV, Network Functions (NFs) can be easily deployed and dynamically allocated. In addition, network resources can be efficiently allocated to Virtual Network Functions (VNFs) through dynamic scaling to achieve Service Function Chaining (SFC)[9] With software-based NFV solutions, some of the NFs are moved to the Service Providers (SPs) to run on a shared infrastructure such as general purpose servers. Therefore, adding, removing or updating a function for all or subset of customers becomes much more manageable since changes could only be done at the ISP rather than at the customer premises as being done today. For SPs, NFV promises to provide the needed flexibility that would enable them to scale up/down services to address changing customer demands, reduce their capital expenditure (CAPEX) and operational expenditure (OPEX) through lower-cost agile network infrastructures, decrease the deployment time of new network services to market. In the context of future 5G networks, NFV ensures optimization of resource provisioning to the end-users with high QoS and guarantee the performance of VNFs operations including minimum latency and failure rate. Essentially, it can ensure the compatibility of VNFs with non-VNFs [113]. To achieve the above benefits, NFV brings three differences on how network services are provisioned compared to traditional practice as stated in [114].

---

[9] Service Function Chaining (SFC) is an ordered list of abstract service functions that should be applied to a packet and/or frames and/or flows selected as a result of classification [112].

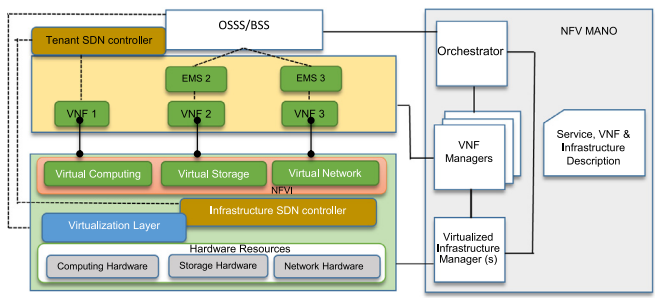

**Fig. 10.** An integration of SDN controllers into the ETSI NFV reference architecture at the two levels required to achieve network slicing.

- *Decoupling of software from hardware platform*: Hardware and software entities in NFV are not integrated, and their functions can progress separately from each other.
- *Greater flexibility for network functions deployment*: Since software are detached from hardware, both software and hardware can perform different functions at various times. This enables operators to deploy new innovative services using the same hardware platform.
- *Dynamic network operation and service provisioning*: Network operators can introduce tailored services based on customer demands by scaling the performance of NFV dynamically.

It is important to note that, while the full-blown software-based implementation using SDN and NFV concepts comes with these benefits, the question is whether the 5G design considerations can meet some technical performance requirements of different verticals needed by Telco Cloud or service providers. A blueprint of the ETSI NFV framework is discussed next.

### 4.3.1. NFV management and orchestration (NFV MANO) framework

The NFV concept in operator infrastructures [115] was first explored by the European Telecommunication Standard Institute (ETSI), mostly to address the challenges towards flexible and agile services and to create a platform for future network monetization. Since then, the NFV reference architecture shown in Fig. 10 was proposed [116] followed by a proof of concept (PoC) [117]. The ETSI MANO framework consists of functional blocks which can be grouped into the following categories: the NFV Infrastructure (NFVI), NFV Management and Orchestration, Network Management System and VNFs and Services. These entities or blocks are connected together using reference points[10] For a complete description of the NFV MANO framework and its entities, we refer the reader to [39,84,114].

Apart from the building blocks of the NFV MANO shown in Fig. 10, the ETSI proposal includes two SDN controllers in the architecture [118]. Each controller centralizes the control plane functionalities and provides a general view of all the connectivity-related components it manages. These controllers are:

- *Infrastructure SDN Controller (ISDNC)*: Provides the required connectivity for communicating the VNFs and its components by managing the underlying networking resources [119]. As managed by the VIM, this controller may change NFV infrastructure behavior on demand according to VIM specifications adapted from tenant requests [12].

---

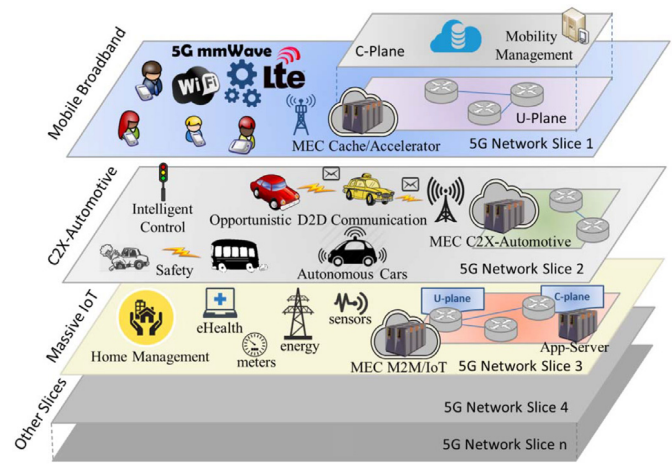

**Fig. 11.** The role of MEC for 5G network slicing.

- *Tenant SDN Controller (TSDNC)*: Dynamically manages the pertinent VNFs, the underlying forwarding plane resources used to realize the tenant's network service(s). The TSDNC is instantiated in the tenant domain [12] as one of the VNFs or as part of the NMS. Note that, both controllers manage and control their underlying resources via programmable southbound interfaces, implementing protocols like OpenFlow, NETCONF, and I2RS.[11] Each controller provides a different level of abstraction. While the TSDNC provides an overlay comprising tenant VNFs that define the network service(s), the ISDNC provides an underlay to support the deployment and connectivity of VNFs [76,92]. For the TSDNC, the network is abstracted in terms of VNFs, without notions of how those VNFs are physically deployed. The ISDNC is neither aware of the number of slices that utilize the VNFs it connects, nor the tenants which operate such slices. Despite their different abstraction levels, both controllers have to coordinate and synchronize their actions in order to achieve the management of network slices on 5G networks [118].

## 4.4. Multi-access edge computing (MEC)

MEC [120] offers application and content providers cloud-computing capabilities and an IT service environment at the edge of the mobile network [121]. MEC processes data close to where it is generated and consumed. This enables the network to deliver ultra-low latency required by business-critical applications and support interactive user experiences in busy venues such as shopping malls and train stations. By processing data locally, MEC applications can also significantly reduce data transfer costs [61].

With this position, MEC results in several essential network improvements, including: (a) enhanced QoS/QoE to end users in case of video streaming enabled through the use of 5G network slicing, (b) optimization of mobile resources by hosting compute-intensive applications at the network edge, and (c) transforming access nodes into intelligent service hubs where context-aware services (e.g., user location, cell load and allocated bandwidth) can be provided with the help of RAN information.

A blueprint of the role played by MEC for 5G network slicing is shown in Fig. 11. As presented by Sciancalepore et al. [122] in a compound architectural evaluation of MEC and NFV, the fundamental element of MEC is the MEC application server, which runs on top of the MEC NFVI infrastructure and provides services to the end-users, implemented as individual MEC Applications (MEC Apps). MEC Apps share communication interfaces with the MEC

---

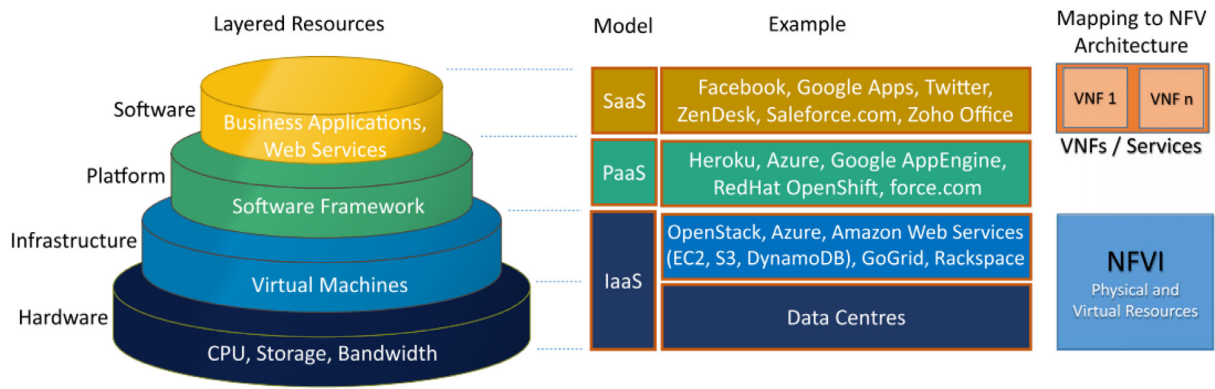

**Fig. 12.** Cloud computing service models and their mapping to part of the NFV reference architecture [84].

platform, where MEC services are hosted. The latter provides services to the Apps and act as an API intermediate between the MEC platform and App. MEC service nodes can operate locally inside the deployed data center or remotely in the cloud. Both MEC Apps and MEC services incorporate interfaces to the Traffic Offload Function (TOF) which is located in the data plane and prioritizes traffic via transparent, policy-based packet monitoring and redirection. This simplifies MECs' integration to the RAN and plays a vital role as a generic monitoring-assisting element [63].

### 4.5. Cloud/fog computing

Cloud computing [123] offers on-demand provisioning of various applications, platforms, and heterogeneous computing infrastructures such as servers, networks, storage, service and applications. According to Mijumbi et al. [84], the traditional role of service provider on a cloud computing environment is divided into two categories namely: (a) the Infrastructure Providers (InPs), and (b) Service Providers (SPs). The InPs manage cloud platforms and lease resources according to a usage-based pricing model while SPs rent resources from one or many InPs to serve the end users. The cloud model consists of three service models [84,123] as shown in Fig. 12 which also indicate their mapping to the NFV reference architecture described in Section 4.3. The service models of cloud computing as defined in [84] include:

- *Software as a Service (SaaS)*: The user can utilize some applications and services running on a cloud infrastructure. A service provider hosts the applications at its data center and a customer can access them via a standard web browser.
- *Platform as a Service (PaaS)*: Provides a platform that allows customers to develop, run, and manage different applications without the complexity of building and maintaining the cloud infrastructure.
- *Infrastructure as a Service (IaaS)*: Provides self-service models for accessing, monitoring, and managing remote data-center infrastructures, such as compute, storage and networking services. Examples of IaaS includes the Amazon Web Services (AWS), Microsoft Azure and Google Compute Engine (GCE)[12].

### 4.6. Network hypervisors

Network hypervisors [124] are the network elements that abstract the physical infrastructure (e.g., communication links, network elements, and control functions) into logically isolated virtual network slices. In physical SDN network, a network hypervisor offers high-level abstractions and APIs that greatly simplify the task of creating complex network services. Moreover, the network hypervisor is capable of inter-networking various SDN providers together under a single interface/abstraction so that applications can establish E2E flows without the need to see or deal with the differences between SDN providers [125]. Through hypervisors, it is possible even to implement higher layer services such as load balancing servers and firewalls or link and network protocol services belonging to L2 and L3 [84]. In the context of network hypervisors [15], the concept of network slicing has been explored in several works such as OpenVirteX [126] and FlowVisor [127], OpenSlice [128], MobileVisor [129], RadioVisor [130] and Hyper-Flex [131]. The MobileVisor [129] can slice the mobile packet core network infrastructure into different virtual networks belonging to different MVNOs. However, most of the network hypervisors (e.g., OpenVirteX and FlowVisor) have been designed for slicing a fixed and wired SDN network. We refer the reader to [15] for a more comprehensive work on network virtualization hypervisors using SDN.

### 4.7. Virtual Machines

Virtual Machine (VM) [132] enables the virtualization of a physical resource where an experimenter can run his/her own Operating System (OS). The basic principle of a VM is that resources such as computing, storage, memory, and network are shared among VMs. However, the entire operational functions of a VM is isolated completely from that of the host and another guest VMs [24,133]. It is also possible to run multiple VMs at the time on one physical machine.

### 4.8. Containers

Containers are light-weight alternatives to hypervisor-based VMs [134] and are created based on the idea of OS-level virtualization. A physical server in containers is virtualized such that standalone applications and services can be instantiated on an isolated servers [24]. Different from VM-based counterparts, containers do not need hardware indirection and run more efficiently on host OS leading to higher application density. Examples of container-based virtualization include: Docker [135], Linux-Vserver [136], OpenVZ [137], and Oracle Solaris Container[13]. In this vein, VMs and containers are capable of running VNFs chained together to deliver a 5G network service or application flexible and therefore forming a base functionality for 5G network slicing. It is important to note that, while containers can efficiently support 5G network slices with highly mobile users, VMs may offer full logical isolation for operating VNFs in a network slice [24].

---

[12] https://apprenda.com/library/paas/iaas-paas-saas-explained-compared/.

[13] https://www.oracle.com/technetwork/server-storage/solaris/containers-169727.html.

**Table 6**
The relationship and comparison between SDN and NFV.

| Category | NFV (Telecom Networks) | SDN (Data Center Networks) | Already Adopted |
|---|---|---|---|
| Network Control | Seamless control and dynamic provisioning of NFs | Provide a centralized network control | Yes |
| Architectural Design | Service or NFs abstractions | Networking Abstractions | Yes |
| Main Advantage | Offering flexibility needed by network | Offering programmable network with open control interfaces | Yes |
| Cost Efficiency | Replace hardware with software | Operational efficiency and energy consumption reduction | Yes |
| Standard Protocol | Supporting multiple control protocols | OpenFlow is the de-factor standard protocol | Yes |
| Leaders/Business Initiator | Born in Telcom Service Providers | Born for networking software and hardware vendors | N/A |
| Formalization | ETSI | ONF | N/A |

### 4.9. Summary and lesson learned

Fig. 7 shows a summary and a tutorial contribution of section 4. To summarize, an achievable step so far in the design patterns of network softwarization has been to identify on how network services and the associated resources that are implemented, according to an SDN architecture, might be integrated within the NFV architectural framework [118]. It is worth stressing that, both SDN and NFV seek to drive a future software-based 5G networking solution that offers a flexible and automated feature selection for network connectivity and QoE provisioning to the end-users. For example, while SDN decouples the control plane from the data/packet forwarding plane, the NFV decouples NFs from dedicated hardware devices.

To this end, SDN and NFV have been accoladed for network softwarization towards 5G systems. Although the two (SDN and NFV) have a lot in common, yet their main difference is that SDN requires a new network platform where the control and data forwarding planes are decoupled. This is not the case with NFV which can run on legacy networks since NFs can reside on commodity servers. We give the relationship and comparison of SDN and NFV in Table 6. We also provide a highlight on the relationship between VMs, cloud computing and NFV using Fig. 12. What remains to be seen from both, the academia and industry is the output of all these technologies toward making 5G network slicing a reality as foreseen and proposed by vendors, operators, and SPs. We discuss next the state-of-the-art of 5G network slicing architectures and their implementations.

## 5. State-of-the-Art: 5G network slicing architectures and implementations

The development of 5G network and its standardization is taking place within several projects and standard bodies. In order to deploy 5G in alignment with market demands, a number of standard bodies (e.g., 3GPP [54], ITU [138], IEEE [139]), associations (e.g., ETSI [63], TIA [140], alliances (e.g., NGMN [45] and Wireless World Research Forum (WWRF) have devoted some initiatives for conducting research and standards on the future mobile networks specifically targeting on 5G of 2020 and beyond. Major telecommunication companies such as Nokia Solutions and Networks [141], Huawei [32], Ericsson [142], ZTE [143], Samsung Electronics [144], Datang [145], Qualcomm [146] and NTT-DOCOMO [147] have already presented and contributed white papers on 5G. The HORIZON 2020[14] and METIS (Mobile and wireless communications Enablers for the Twenty-twenty (2020) Information Society) are the major 5G research projects initiated and funded by the European Union (EU) [2,150,151].

[14] A complete list of other 5G-PPP Phase I and II Projects which are funded by the EU under H2020 can be found in [148] and [149].

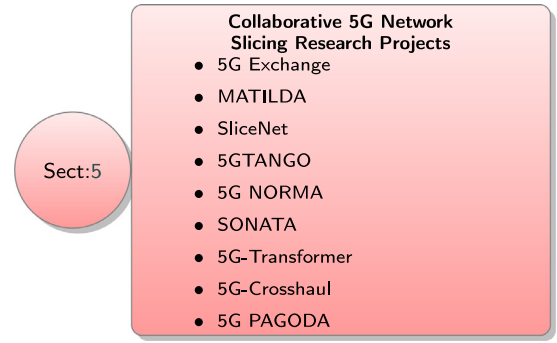

**Fig. 13.** Summary of 5G network slicing projects, architectures and implementations in Section 5.

The common goal and vision of industrial and research perspective have been to design 5G as a network that can meet the requirements of different verticals while satisfying the end-users' service quality demands. For example, focusing on QoE management in the future 5G architecture, the PPP FP7 FIWARE [152], 5G-NORMA [153,154] and MIUR PLATINO project [155] have been working towards the realization of orchestration algorithms for control decisions, and different mechanisms for subjective QoE personalization/differentiation and the end-users' QoE. Projects such as 5G-Xhaul [151] and SELFNET [156] have been initiated to realize self-healing, self-configuration and self-optimization capabilities for 5G networks. As the NGMN continues to work on 5G network slicing concept, several other standards organizations (e.g., ETSI, ITU-T, 3GPP), academia and industrial research projects (5G-NORMA, 5GEX) and vendors are working in parallel with different objectives, and some of them in close collaboration with the ETSI. In this section, we explore 5G network slicing research projects in terms of their architectures and different implementation details. Fig. 13 shows a summary of collaborative 5G network slicing research projects.

### 5.1. Collaborative 5G network slicing research projects

#### 5.1.1. 5G Exchange (5GEx)

5GEx [157] is set to provide a multi-operator collaborative approach by developing an SDN/NFV based multi-domain, multiservice orchestration platform to provide services *"manufactured by software"* on 5G networks. 5GEx will allow E2E network and service elements to be combined and operate together in multivendor and resource 5G virtualized environments. From the technical perspective of orchestrating resources on 5G systems, the developed architecture is to ensure that both network resources and slices are provided on demand-basis. To summarize, 5GEx's objectives are: (1) to develop a multi-domain and multi-service infrastructure for 5G networks based on SDN/NFV, (b) enable orchestration of services and an IaaS model for multiple carriers forming the so-called *"5G network factory,"* [158].

As pointed out by Sgambelluri et al. [158], apart from catering to the needs of future 5G services, 5GEx is positioned to overcome also the historical, technological and market fragmentation of the European telecommunications sector. Such a generic, open, and standardized offering of various connectivity modes supported with other 5G capabilities will enable the numerous small to medium-sized enterprises (SMEs) and content providers to differentiate and monetize their online content and application provision [159]. It is intuitively mentioning that the core element in the 5GEx infrastructure is a slice that efficiently serves 5G verticals by relying on lower-level 5GEx basic services and SDN/NFV techniques. Standard interfaces are used to connect and exchange information among entities as shown in Fig. 14. The multi-domain orchestrator interface (1) is used to translate 5GEx service requests from customers to a chain of VNFs with their associated resource requirements. Interface 2 trade slices inline with the ELAs/ELAs and 5GEx higher-level services among 5GEx-enabled orchestrator. Interface 3 is responsible for the management of own or leased resources through interface 2.

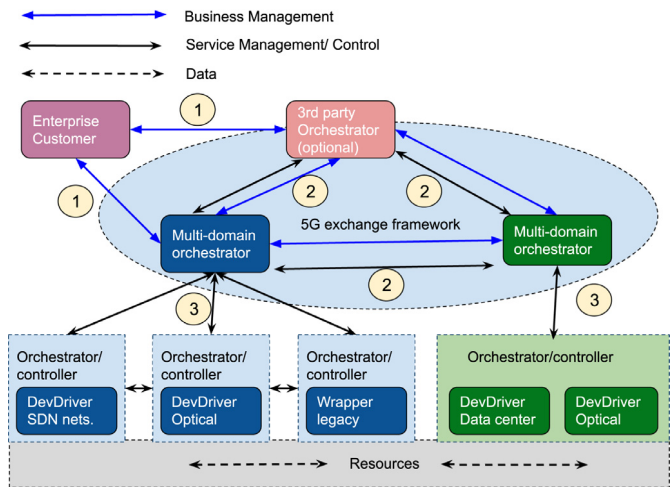

**Fig. 14.** 5GEx network slicing conceptual architecture.

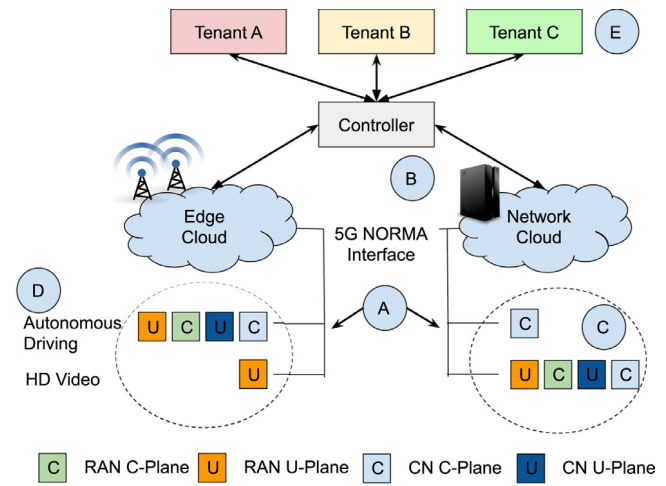

**Fig. 15.** The main innovations of 5G NORMA concept.

It is worth noting that, precursor projects containing ideas and code for 5GEx interfaces include the UNIFY (interface 3, [160]), and T-NOVA (interface 1, [161]). The 5Gex framework supports a variety of collaborative models such as the *"Direct peering"* for distributed multi-party collaboration. It also supports higher-level abstractions and advanced models covering views, resources, and services across several exchange points or points of presence (PoPs). The customer-facing *"3rd party orchestrator"* in Fig. 14 refers to a virtual mobile network operator who implements the multi-domain orchestrator functionality but does not own an infrastructure.

### 5.1.2. MATILDA

MATILDA [162] aims to design and implement a holistic 5G E2E services operational framework that solves the orchestration of 5G-ready applications and services over sliced programmable infrastructure [162]. Smart and unified orchestration strategies are applied for creating and maintaining the required network slices [163]. Cloud/edge computing and IoT-based resources are mainly supported by a multi-site virtualized infrastructure manager, while the lifecycle management of the supported VNF Forwarding Graphs (VNF-FGs) and a set of network management activities are provided by a multi-site NFV Orchestrator (NFVO).

### 5.1.3. SliceNet

SliceNet aims at maximizing the potential of infrastructure sharing across multiple operator domains in SDN/NFV-enabled 5G networks [164]. The project intends further to achieve a genuinely

E2E network slicing through a highly innovative slice provisioning, control, management and orchestration mechanisms which are QoE oriented to 5G verticals [165]. SliceNet aims towards the maximization of network resources sharing within and across different administrative domains. That way, Slicenet is to create and form a close partnership between industry and vertical business sectors in achieving the fully connected society vision in 5G [166]. Building on these objectives, SliceNet covers three vertical use-cases, namely, (1) 5G smart grid self-healing, (2) 5G smart m-health, and (3) 5G smart city [167].

### 5.1.4. 5GTANGO

5GTANGO [168] addresses significant challenges associated with both the development and deployment of the complex services envisioned for 5G networks. The core objective of 5GTANGO [168] is to develop an extended DevOps model that accelerates the NFV uptake in the industry at a scale of network service capabilities of the 5G platform in vertical showcases [169]. To date, a general 5G architecture for multi-site NFVI PoP that supports network slicing and multi-tenancy has been presented in [169] while a network slicing resource allocation and monitoring framework over multiple clouds and networks called *"Netslice planner"* is demonstrated in [170]. Kapassa et al. [171] present an automated proposition and management mechanisms for enforcing QoS/QoE agreements. Kapassa et al. [172] further propose a framework that facilitates the VNF and network slices- tailored SLAs management in 5G. In that aspect, 5GTANGO puts forth the flexible programmability of 5G networks with a modular service platform having an innovative orchestrator in order to bridge the gap between business needs and network operational management systems [173].

### 5.1.5. 5G NORMA

5G NORMA [153] proposes a multi-service and multi-tenant capable 5G system architecture based on the concept of network slicing [174]. The transition from legacy to the 5G NORMA system architecture builds on two enablers, namely (a) adaptive decomposition and allocation of network functions using a Software-Defined Mobile network Orchestration (SDMO) and (b) network programmability via a Software-Defined Mobile Network Control (SDMC). Fig. 15 shows the fundamental entities of the 5G NORMA architecture including [153]: (1) the *Edge Cloud* which is composed of the bases stations and the remote controllers that are deployed at the radio or aggregation sites, (2) the *Network Cloud*, one or more data-centers that are deployed at central sites, and (3) the *Controller* that organizes and executes the NFs which are co-located in the network cloud.

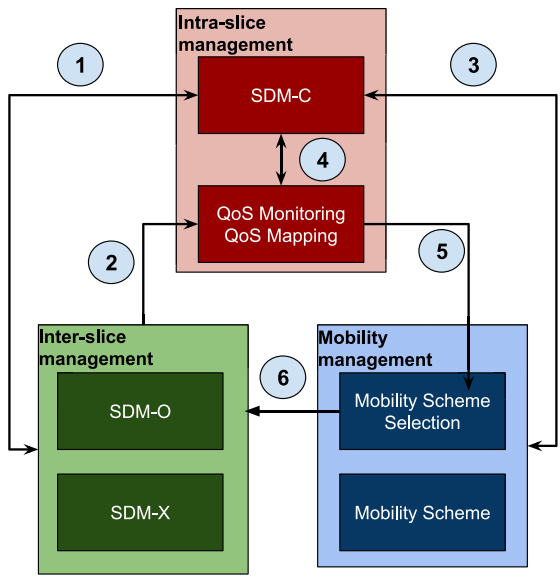

**Fig. 16.** 5G NORMA architecture building blocks and their interactions.

Fig. 15 also illustrates five main pillars (A to E) and three innovative functionalities of 5G NORMA architecture. The five pillars include Pillar A that indicates the adaptive decomposition and allocation of mobile NFs between the edge and network cloud, basically depending on the deployment needs and service requirements. Pillar B indicates the SDMO which use the principles of SDN to perform mobile network specific functions. Pillar C signifies the joint optimization of both mobile access and core NFs localized together, either in the edge cloud or the network cloud. The two innovative aspects of 5G NORMA functionalities are included in pillar D to provide multi-service and context-aware adaptation of NFs as well as supporting a variety of services and their corresponding QoS/QoE requirements. Finally, pillar E highlights the mobile network multi-tenancy that supports the on-demand allocation of radio and core resources towards virtual operators and vertical market players. One of the key strengths and the spirit of 5G NORMA is *softwarization* described in Section 3.1 that provides the needed flexibility in the implementation of mobile NFs other than routing and forwarding. The 5G NORMA provides flexible connectivity of 5G networks using six building blocks: the Software Defined Mobile Network Controller (SDM-C), Orchestrator (SDM-O), Coordinator (SDM-X), the QoE/QoS Mapping and Monitoring module and the Mobility Management module.

Fig. 16 shows an overview of 5G NORMA functional blocks as well as the interactions among them. The numbers in Fig. 16 indicate the following entities: (1) resource pool management (2) resource requests (3) service requirement extractions (4) mobility information feed (5) mobility-driven orchestration (6) mobility requirements and (7) QoE/QoS reporting. We next describe each of the building blocks as follows:

- *SDM-C*: Is set to enable flexible network management and operation within a network slice. It specifies both northbound and southbound interfaces which enable different functionality [175]. As such, the northbound interface is used to control network operation in terms of QoE/QoS and mobility management, whereas the southbound interface conveys the required actions within a given network slice. The SDM-C receives the network requirements through the northbound interface and, once processed, triggers the necessary operations through the southbound interface [176].

- *QoE/QoS Mapping and Monitoring*: Enables the monitoring of QoE/QoS parameters within a network slice and therefore allowing the SDM-O to act accordingly in order to fulfill the network requirements and the agreed ELAs/SLAs. It further allows allocating the minimal amount of resources for achieving the required QoE which in turn avoids user's churn and improves energy efficiency [177].

- *SDM-X*: Enables the control of shared NFs or resources among selected network slices. It receives information from the SDM-O block and process them so that it can decide whether shared resources among network slices upon a request coming from SDM-C can be modified or not. The SDM-X is also responsible for controlling VNFs/PNFs in a common 5G network data and control layer. As such, it needs to ensure the fulfillment of the received requirements within its corresponding network slice [178].

- *SDM-O*: Enables the support of multi-service and multi-tenancy using network slicing to orchestrate resources between slices belonging to different administrative domains. The SDM-O analyzes service requests and feeds the results to the network slice creation life-cycle. The SDM-O is further broken down into Service Orchestration, Slice Orchestration, and Inter-slice/Inter-tenant Orchestration. SDM-O has complete knowledge of the network and is responsible for managing resources needed by all tenants' slices. That way, it enables the orchestrator to perform the required optimal configuration in order to adjust the number of used resources and, hence, making efficient use of the network resources [153,174]. Fig. 17 shows the life-cycle of a network slice creation and operation based on the 5G NORMA architecture. Similar to the process of the IETF Service Function Chaining (SFC) WG4 [179], the SDM-O maps the general service requirements in terms of KPIs (e.g., SLAs) to requirements that are used to build the actual chain of VNF, starting from a template library. It is worth mentioning that, SDM-O handles slices creation requests associated with a well-defined service (e.g., Vehicular, IoT), possibly those belonging to different tenants.

- *Mobility management*: This block is implemented as an SDM-C application that collects information from the QoE/QoS module and enforces new rules through the SDM-C southbound interfaces. Two sub-modules, namely, the *mobility management scheme selection* and the *mobility management scheme design* are considered in the mobility management component. The latter includes all the algorithms needed to perform a certain mobility management functions, while the former performs the selection of the most appropriate slice design based on the slice requirements [153,176].

*5.1.6. SONATA*

SONATA aims at increasing the flexible programmability of the 5G network by developing (a) a novel Service Development Kit (SDK) [181], and (b) a modular Service Platform & Orchestrator (SPO) [182]. Intuitively, four innovations in the SONATA system stand out among NFV MANO platforms namely: (1) the modular and customizable MANO architecture as proposed in [183] that provides flexibility to network operators and ability to add new features via plug-ins, (2) an interoperable and vendor agnostic framework to provide a multi-VIM, multi-vendor and multi-site capabilities on the underlying ETSI-based architecture, (3) an efficient network service development and NFV DevOps [184] that provide service developers with a SDK for efficient creation, deployment and management of VNF-based network services, and (4) the 5G *slicing support* and *recursion support*. The slicing support in SONATA is set to deliver performance isolation and bespoke network configuration for industry verticals that are foreseen in 5G networks while the recursion support allows stacked tenant and large-scale deployments in new software network business models [185].

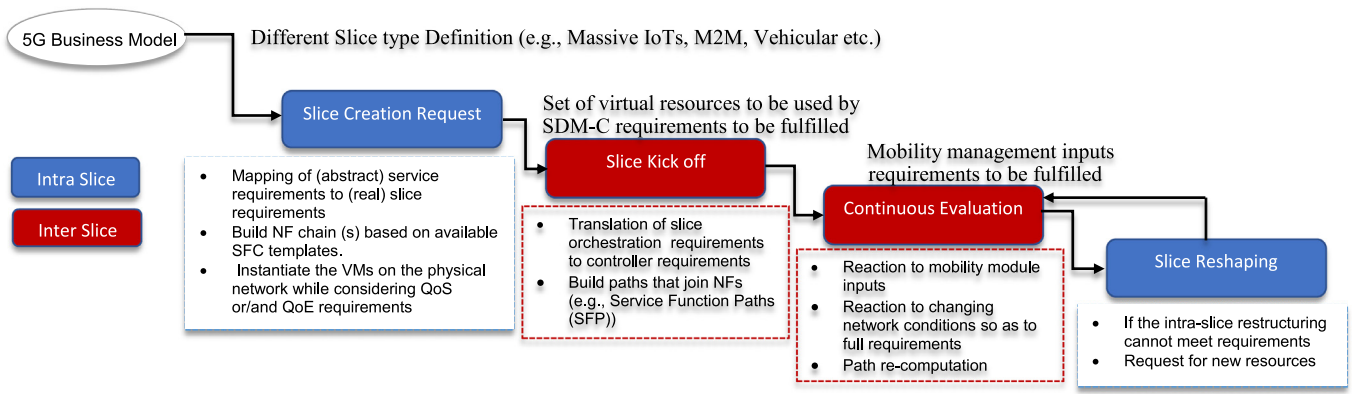

**Fig. 17.** The life-cycle of a network slice in 5G NORMA architecture.

### 5.1.7. 5G-MoNArch

5G-MoNArch leverages the concept of network slicing to design and develop a flexible, adaptable, and programmable 5G architecture that will support a variety of use cases in vertical industries such as automotive, healthcare, and media [186]. The project evolves and enhances the concepts from 5G NORMA [174] and METIS II[15] to a fully-fledged architecture and develop prototype implementations that can be applied to typical use cases. In order to achieve this, three novel innovations are explored, namely: (a) inter-slice control and cross-domain management, specifically to enable the coordination across slices and domains, (b) experiment - driven optimization 5G performing algorithms, and (c) cloud-enabled protocol stack that provides the needed flexibility in the orchestration of VNFs. One of the key element in the 5G-MoNArch architecture is the M & O layer which complies with 3GPP specifications for the management and orchestration of network resources [187]. The current 5G-MoNArch architecture [188] explicitly takes into account the interaction with the 3GPP management entities dedicated to network management.

### 5.1.8. 5G-Transformer

5G-Transformer aims to transform today's mobile transport network into an SDN/NFV-based Mobile Transport and Computing Platform (MTCP) to manage slices tailored to the specific needs of vertical industries [189]. In this aspect, 5G-Transformer is set to deliver a scalable MTCP by adding the support of (a) an integrated MEC services, (b) dynamic placement and migration mechanisms of VNFs, (c) new mechanisms for sharing VNFs by multiple tenants and slices, (d) new abstraction models for vertical services [190], and (e) customized profiles for the C-RAN functional split considering the requirements from different verticals. To date, SDN control solution for automatic operations and management of services in a fixed-mobile converged packet-optical 5G network is proposed in [191] while the techno-economic solution for future software-defined 5G converged access networks are highlighted in [192]. Aissioui et al. [193] introduce the Follow Me edge-Cloud (FMeC) concept that leverages the MEC architecture to sustain the 5G automotive system requirements. The FMeC is to ensure low-latency access to automotive services and applications that are deployed at the edge-cloud in the context of 5G.

### 5.1.9. 5G-Crosshaul

Future 5G networks will require fronthaul and backhaul solutions between the RAN and the packet core networks [4]. As an attempt to address this, the 5G-Crosshaul has devoted efforts to develop a flexible and software-defined 5G integrated backhaul and fronthaul transport network where reconfiguration of network elements is done in a multi-tenant environment [196,202]. The 5G-Crosshaul architecture implementation consists of: (i) a control plane that provides an abstraction of a network model and integrates the Xhaul Control Infrastructure (XCI); (ii) a unified data plane to provide novel Xhaul Packet Forwarding Element (XFE). It is worth highlighting that, 5G-Crosshaul is to enable network slicing as a service that addresses the dynamic allocation of slices over a shared softwarized infrastructure [93]. Such allocation of slice involves the selection of NFs, their constrained placement, and the composition as well as configuration of the underlying physical/virtual infrastructures. That way, it fulfills the 5G services' requirements for example, in terms of latency, bandwidth, and processing capacity. Key network slicing services considered to enable explicit control, automation and slice management include: (a) the provisioning of a tenant's owned network services similar to those defined by the ETSI NFV architecture [114], and (b) virtual infrastructures (VIs) [203]. This goes along with an IaaS model that enables provisioning of 5G services under the control and operation of different tenants. The VIs deployment is oriented to the business-to-business (B2B) market, targeting customers like MVNOs and cloud providers specializing in customizable IaaS services. Conversely, network services target customers operating in the B2C segment such as application/service providers (e.g., multimedia content providers) that offer streaming services to end users.

### 5.1.10. 5G!PAGODA

5G!PAGODA [81] represents the next evolution step in softwarized networks through the development of a scalable 5G slicing architecture that supports network slices composed on multi-vendor NFs [76,81]. The 5G!PAGODA architecture is aimed at providing efficient network slice management and orchestration mechanisms in distributed, edge dominated network infrastructures through a lightweight control plane and data plane programmability [24]. It is worth stressing that, the proposed 5G!PAGODA architecture takes the concept of mobile networking to the next level such that slices of Virtual Mobile Networks (VMN) are created on-demand basis and customized according to the changing needs of mobile services using physical resources across multiple-domains [76], an approach that leverages the ETSI NFV architecture. From multi-slice management system perspective, the ETSI NFV MANO is extended with three functional modules, namely (a) a *Multi-Domain Slice Orchestrator (MDSO)*, (b) *Business Service Slice Orchestrator (BSSO)* and (c) the *Domain -Specific slice Orchestration (DSSO)*. In this article we only give highlights on these modules since the functionality of all other entities are similar to those stipulated in the ETSI MANO framework (see in [114]).

---

The MDSO provides a slice on top of multiple administrative domains. It announces and informs the tenant through the MSSO and/or the slice-specific OSS of any ELAs/SLAs breaches or any other types of major failures of the deployed slice. The MDSO also implements a Slice Placement Function (SPF) that allocates and interconnect slice-specific VNFs according to service requirements (e.g., latency) and other network resource constraints. The DSSO receives information regarding the life-cycle management of network slices from the multi-domain slice orchestrator. It communicates and transfer these information to multiple NFVOs that are within the same administrative domain. The BSSO is responsible to advertise the available services and reconfigure tenant's slices after their deployment. As such, it provides the API for tenants and different verticals to query network resources availability and slice pricing information. The BSSO provides capabilities for tenants or verticals to destroy slices and deploying new ones [204,205].

### 5.1.11. NECOS

NECOS [206] builds on the concept of Cloud Slicing (CS)[16] to propose the Lightweight Slice Defined Cloud (LSDC) solution, an approach set to achieve the process of optimal cloud configuration automatically. The NECOS extends the virtualization concept described in Section 4.5 to all resources which spans multiple cloud infrastructures, from the data center to the edge [206]. The key novel aspects of the LSDC are to: (a) present a new *SaaS* model, (b) enable the configuration of slices across physical resources in the cloud to better accommodate various 5G service demands, (c) allow each tenant that comprises the cloud environment to be managed via software, and (d) utilize lightweight and uniform management systems, with small footprint components, deployable on a large number of small servers and cloud systems [204].

In light with the concept of CS, the NECOS architecture shown in Fig. 19 adopts elements from other 5GPPP EU project architectures (e.g., the SONATA [182], the 5GEx, and [158]) to build a unified environment that integrates connectivity, computation, and storage in order to create the SaaS model. The NECOS platform exposes interfaces for both service deployment and resource allocation using several modules such as the *Network Manager, Cloud Manager, Control Element for VMs* and the *Service Orchestration* within a single deployable and distributed 5G infrastructures. The tenant of NECOS can be a CDN company that requires slices for running their services. The LSDC or the NECOS Slice Provider is the component that enables the creation of full E2E slices from a set of fundamental slice parts [205]. The LSDC indicates a northbound API that is compatible with a tenant's service orchestrator. That way, it enables tenants either to operate on the full infrastructure or to choose to interact with SaaS providers. The Slice Resource Orchestrator (SRO) is the component responsible for the orchestration and management of slices at the run-time of their lifecycle. The SRO is also responsible for embedding and the actual placement of VMs as well as virtual links for the services into different resource domains.

The Slice Builder within the LSDC is responsible for building a full E2E multi-domain slice from the relevant constituent slice parts by searching resources that are available from the marketplace. This resource lookup involves continuous communication with a Slice Broker, an entity responsible for contacting Slice Agents. As shown in Fig. 19 each Slice Broker have access to many Slice Agents from different geographical domains, which can provide offers for the required slice parts that match a set of request constraints. To interact with the actual remote cloud ele-ments, the LSDC uses an Infrastructure and Monitoring Abstraction (IMA) mechanism which allows the Slice Provider to interact with various remote VIMs, WIMs, and monitoring sub-systems in a generic way using plug-in adaptors with the relevant API interactions [206]. It is worth mentioning that, the Resource Providers represent organizations such as data centers that can provide resources in the form of servers, storage, and network, etc. To aid in flexibility, the Resource Marketplace (e.g., telecoms) provides the way for the NECOS Slice Provider to find the slice parts to build up a slice [205].

### 5.2. Summary and lesson learned

This section provides various collaborative 5G network slicing research projects. We note that multi-service and multi-tenancy is a concept that has been addressed in many research 5G slicing projects. In the scope of 5GPPP, projects like 5G-NORMA and SESAME are addressing RAN multi-tenancy [85] while CHARISMA covers 5G access networks. The 5G-Crosshaul complements with these efforts by focusing on the transport network aspects directly related to the combined fronthaul and backhaul, targeting per-tenant services that combine computing, storage, switching and transmission resource management. Table 7 presents a summary of academia/industry 5G projects in terms of their focus area, QoS/QoE support and their SDN/NFV related works. Table 8 also provides comparison summary of different 5G architectural approaches in terms of practical implementation, technology adoption and deployment strategy. These 5G collective efforts will enable cross-domain orchestration of services over multiple administration multi-domain. It will also allow the context-aware adaptation of NFs to support a variety of services and their corresponding QoE/QoS requirements.

## 6. Open source orchestrators, proof of concepts & standardization efforts

### 6.1. Open source orchestrators for network slicing

Orchestrators for 5G network slicing are becoming complementary to allow fast innovation, which is why most of the current solutions are open-source [207]. In order to realize the network slicing vision, an *Orchestrator*, a software responsible for automating the creation, monitoring and deployment of resources and services in the underlying softwarized and the virtualized environment is required. The ETSI defines two different types of orchestrators [208]: (a) Resource Orchestrator (RO) that coordinates, authorizes, releases and engages NFVI resources among different Point of Presence (PoPs) or within one PoP, and (b) Service Orchestrator (SO) - that creates E2E service between different VNFs). To date, many open source orchestrators have been developed to realize the dynamic network slices management and orchestration of resources in 5G networks. The following subsections provide a comprehensive survey of orchestrators of which some of them are currently used for realizing the 5G slicing network concept.

### 6.1.1. OSM

An open source management and orchestration (OSM) stack has been developed in accordance with the ETSI NFV information models [209]. The OSM includes the SO, RO and a configuration manager and targets the requirements of commercial NFV networks. Using the OSM orchestrator, an automated assurance and DevOps in service chains and 5G network slices are demonstrated in [210]. As an example, Fig. 20 shows three slices with different QoS requirements running on a network that spans several elements. Throughput is the KPI for a mobile broadband slice that is required for residential subscribers to assure their SLA/ELAs requirement.

---

[16] Cloud Slicing provides concurrent deployment of multiple logical, self-contained, independent, shared or partitioned slices on a common infrastructure platform. CS also enables dynamic multi-service and multi-tenancy support for 5G vertical market players.

**Table 7**
A summary of academia/industry 5G projects and implementation based on SDN/NFV.

| Name | Focus Area | | QoE | SDN/NFV Related Work |
|------|------------|------|-----|----------------------|
| | SDN | NFV | | |
| 5G-NORMA [153] | Yes | Yes | Yes | Multi-service and context-aware adaptation of network functions to support a variety of services and corresponding QoE/QoS requirements. |
| 5G-MEDIA | Yes | Yes | Yes | A flexible network architecture that provides dynamic and flexible UHD (4K/8K) content distribution over 5G CDNs |
| 5G-MoNArch | Yes | Yes | Yes | Employ network slicing to support the orchestration of both access and core network functions, and analytics, to support a variety of use cases in vertical industries such as automotive, healthcare, and media. |
| 5GTANGO | Yes | Yes | Yes | To develop a flexible 5G programmable network with an NFV-enabled Service Development Kit (SDK) that supports the creation and composition of VNFs and application elements as "Network Services". |
| SESSAME | Yes | Yes | Yes | Develop programmable 5G network infrastructure that support multi-tenancy, decrease network management OPEX whilst increasing the QoS/QoE and security. |
| MATILDA | Yes | Yes | No | Orchestration of 5G-ready applications and network services over sliced programmable platforms. |
| 5G-Transformer | Yes | Yes | No | Develop an SDN/NFV-based 5G network architecture that meet specific vertical industries' (e.g., eHealth, automotive, industry 4.0 and media) requirements. |
| 5G-Crosshaul [93] | Yes | Yes | Yes | The design of 5G transport architectural solution that supports multi-domain orchestration among multiple network operators or service providers (e.g.,, multiple tenants). |
| 5G-XHaul | Yes | Yes | Yes | Develop a scalable SDN control plane and mobility aware demand prediction models for optical/wireless 5G networks. |
| CogNET [180] | Yes | Yes | No | Dynamic adaptation of network resources of VNFs, whilst minimizing performance degradations to fulfill SLA/ELAs requirements. |
| CHARISMA | Yes | Yes | Yes | To develop a software-defined converged fixed 5G mobile network architecture that offers both, multi-technology and multi-operator features. |
| SaT5G | Yes | Yes | Yes | Integrated management and orchestration of network slices in 5G SDN/NFV based satellite networks. |
| SLICENET | Yes | Yes | Yes | Develop a cognitive network control, management and orchestration framework, that supports infrastructure sharing across multiple operator domains in SDN/NFV-enabled 5G networks. |
| SONATA | Yes | Yes | Yes | Enable an integrated management and control to be part of the dynamic design of the softwarized 5G network architecture. |
| COHERENT | Yes | Yes | No | Efficient radio resource modelling and management in programmable radio access networks. |
| 5G Exchange [158] | Yes | Yes | No | Enabling cross-domain orchestration of services over multiple administrations or over multi-domain single administrations. |

**Table 8**
A qualitative comparison of orchestrators and deployment options [201] .

| Reference | Network Slicing | Evaluation Methodology | Deployment Strategy | Main Objectives |
|---|---|---|---|---|
| [194] | − | Prototype and simulation | Control and data plane | Optimal use of network resources. |
| [157] | ✓ | Prototype | control and data plane | Realization of network slices. |
| [158] | ✓ | Prototype and simulation | Control and data plane | Provide automated network services across multiple operators. |
| [195] | − | Prototype and simulation | Control and data plane | Optimization of 5G network load costs. |
| [196] | ✓ | Prototype | Control and data plane | Realization of network slices. |
| [197] | − | Prototype and simulation | Data plane | 5G Radio access trials over NOMA channels. |
| [198] | − | Prototype and simulation | Control and data plane | Multi-domain orchestration of services. |
| [12] | ✓ | Prototype | Control and data plane | Realization of network slices. |
| [199] | − | Prototype | Control and data plane | To provide a scalable, flexible and resilient 5G network architecture. |
| [200] | ✓ | Prototype and simulation | Control plane | Provisioning of network slices to MNOs. |
| [13] | ✓ | Prototype | data plane | Realization of NS in RANs. |
| [153] | ✓ | Prototype | Control and data plane | QoE management in 5G. |
| [174] | ✓ | Prototype | Control and data plane | Network slice management. |
| [93] | ✓ | Prototype | Control plane | Network slice management. |

For [197], only a PoC for 5G radio access is given using Non-Orthogonal Multiple Access (NOMA) experiments. Note that, neither SDN nor NFV is considered in their implementation.

Some of VNFs for this network service may include vCache and vDPI.

### 6.1.2. OpenMANO

OpenMANO [211] is an open source project that provides a practical realization of the Management and Orchestration reference architecture (NFV MANO), currently under the ETSI's NFV ISG standardization. The OpenMANO address aspects related to performance and portability by applying Enhanced Platform Awareness (EPA)[17] principles. While encouraging the industry and software developers to explore new NFV possibilities, the OpenMANO provides three software module namely: Openmano, Openvim and the Openmano-GUI. The Openvim is a lightweight that support for the high and predictable performance of the NFV-specific VIM implementation. It directly interfaces with the compute and storage nodes in the NFVI and OpenFlow controller to provide computing and networking capabilities. It offers an OpenStack-like northbound interface (openvim API), where enhanced cloud services are offered including the creation, deletion, and management of instances and networks. It is worth noting that, the OpenMANO is directly provided with a background of NFV for 5G networks. Also, the OpenMANO has a northbound interface (openmano-API), based on REST, where MANO services are offered including the creation and deletion of VNF templates, VNF instances, network service templates, and instances [84].

### 6.1.3. OpenNFV

OpenNFV [212] is a platform developed by HP to facilitate the development and evolution of NFV components and SDN infrastructure across various open source ecosystems. Based on the ETSI NFV reference architecture, the OpenNFV consists of three parts, namely NFV director, NFV manager and OpenStack (HPE Helion). The NFV director performs operations regarding automatic deployment and monitoring of the VNF ecosystem. The NFV director also enables virtualization environments that can efficiently deploy VNF instances while supporting heterogeneous hardware platforms. The NFV manager is responsible for the life cycle management of the VNF instances and enabling scale-up or scale-down of these instances accordingly. The Helion OpenStack offers an open source platform supporting VNFs. To date, the Virtual Central Office (VCO)[18] is one of the use cases supported by OpenNFV to provide a slice of mobile infrastructure for VNO or a tenant.

### 6.1.4. CloudNFV

CloudNFV [213] is an open source platform for implementing NFV- based on cloud computing and SDN in a multi-vendor environment. CloudNFV consists of thee components, namely: the *orchestrator, manager* and an *active virtualization*. The orchestrator addresses the VNF location for a particular service and the connectivity between them. The manager operates on existing resources and maintains an information base of the running services. The NFs and services are all represented by active virtualization using the active resource and active contract sub-elements. It is important to note that, the active contract defines service templates according to the characteristics of the available NFs whereas the existing resource represents the status of infrastructure resources [213].

### 6.1.5. OpenBaton

The OpenBaton [121] is set to improve the NFV performance and grant the security of the overall infrastructure by integrating the underlying software and hardware architectures, networking, management, and orchestration. In essence, it ensures the development of virtual network infrastructures by adapting NFs to the specific cloud environment. The OpenBaton integrates two different engines: (a) the auto scaling engine for managing scaling operations, and (b) the event management engine for dispatching network functions. It is important to mention that, in order to deploy Juju Charms[19] the OpenBaton considers a generic VNFM for the life cycle management of the VNFs based on the corresponding descriptors and a Juju VNFM Adapter. Juju Charms offer an amazing experience for private and public cloud deployments, including bare metal with MAAS, OpenStack, AWS, Azure, Google Cloud and more.

### 6.1.6. Cloudify

Cloudify [214] is a cloud orchestrator based on TOSCA that provides compute, network and storage resources. It provides a complete solution for automating and managing applications deployment and DevOps processes on top of a multi-cloud environment using the IaaS API. That way, Cloudify enables an automatic reaction to pre-defined events with the appropriate corrective measures. It also eliminate the boundaries between orchestration and monitoring of NFs. Cloudify reduces multi-vendor lock-in by offering interoperability among various cloud platforms such as VMware, Cloudstack, Amazon, and Azure. Hess [215] demonstrates the 5G network slicing implementation where vEPC services are deployed over multiple containerized OpenStack clouds, and E2E orchestration of each network slice is performed.

---

[17] https://wiki.openstack.org/wiki/Enhanced-platform-awareness-pcie.
[18] https://www.opnfv.org/resources/virtual-central-office.
[19] https://cloudbase.it/juju/.

### 6.1.7. T-NOVA

T-NOVA [161] leverages the benefits of cloud management architectures and SDN to enable automated provisioning, monitoring, configuration, and efficient operations of Network Function-as-a-Service (NFaaS) on top of virtualized 5G network infrastructure. Following the same principle from the ETSI NFV architecture, the VIM and NFVI are also separated in the T-NOVA design based on OpenStack and OpenDaylight. The T-NOVA consists of two components namely, (a) the Virtualized Resource Orchestrator (VRO) responsible for managing to compute, storage and network resources, and (b) the Network Service Orchestrator (NSO) which maintains the lifecycle of the network services connectivity. The T-NOVA have an additional marketplace layer on top of the orchestrator for implementing business-related functionalities in a multi-user setting, employing the paradigm of "APP-store". The customer-facing module on this layer can allow operators to offer their infrastructures as a value-added service [221].

### 6.1.8. OPNFV

OPNFV [222] is an open source platform that facilitates the development and evolution of multi-vendor NFV components. The OPNFV ensures certain performance targets and interoperability by accelerating the development of emerging NFV products and services. That way, the OPNFV work focuses on particular NFV use cases to conduct performance and use case-based testing on current standard specifications. Note that, the OPNFV takes into consideration components from ONOS, OpenDaylight, OpenStack, KVM, Open vSwitch, DPDK and Linux to mainly concentrates on NFVI and VIM [222].

### 6.1.9. ExperiaSphere

ExperiaSphere [223] revolves around the concept of flexible 5G service models to provide abstractions of the available resources in software-defined and virtualized infrastructures. The management and orchestration of the cloud, SDN, NFV, and even legacy networks resources in ExperiaSphere are formed on a universal service-layer using TOSCA and the Universal Service Definition Language (USDL) principles. TOSCA and USDL define the ExperiaSphere structured intelligence that links data models to service events and the derived operations of virtual network elements.

### 6.1.10. M-CORD

The Mobile - Central Office Re-architected as a Data center (M-CORD) [224] is an innovative solution that leverages the pillars of SDN, NFV and cloud technologies to disaggregate and virtualize RAN and core functions of 5G mobile wireless networks. The main objectives of M-CORD are to (a) provide customized services and better QoE to customers by offering a reduced latency and increased throughput, (b) enhance resource utilization by exploiting real-time resource management, and (c) an agile and cost-efficient deployment of innovative 5G applications and services. M-CORD brings data-center economics and cloud agility to operator's networks. That way, M-CORD lays the foundation for 5G networks to enable the creation of use case-specific services that can be dynamically scaled via a single SDN control plane using ONOS [103] to control the virtual network infrastructure resources [225]. In [226], authors propose a M-CORD-based MEC - enabled architecture for traffic offloading that brings the computation to the proximity of the user in 5G networks. The traffic offloading approach is incorporated to minimize the latency and the load of the core network. It also enables content providers to provide context aware 5G services to the end-users using the collected RAN information. Abbas et al. [227] exploit the M-CORD architecture to propose network management mechanisms for slicing the transport network, core network and the virtualized broadband base unit (vBBU). The slice management application running over ONOS can manage and associates the network slices to user equipment when a service request is received. It is importnat to mention that, M-CORD transforms the 5G mobile network such that the SDN control plane is logically centralized where specific services offered by mobile operators are monitored and scaled dynamically [228].

### 6.1.11. ZOOM

Zero-time Orchestration, Operations and Management (ZOOM) [229] is a TM Forum project that facilitates the development of virtualization and NFV/SDN best practices and standards. It is aimed at identifying new security mechanisms that will protect NFVI. It also defines an operations environment necessary to enable the delivery and management of VNFs. Currently, the ongoing work under ZOOM is divided into several collaborative project areas including: (a) the hybrid infrastructure management platform, (b) network resource lifecycle management, (c) the operations center of the future, and (d) catalysts proof-of-concept projects. Within the context of the catalyst project, ZOOM has been providing demos supported by operators and vendors that establishes DevOps, NetOps and ServOps user scenarios [201].

### 6.1.12. NGSON

The Next generation service overlay networks (NGSON) [230] is the official name related to standardization effort under IEEE Project 1903[20]. NGSON is meant to enable service/content providers, network operators, and end-users to provide and consume composite services by the deployment of self-organizing, context-aware, and adaptive 5G networking capabilities [230]. To support these capabilities, the NGSON functional architecture document [231] specifies a set of functional entities and relationships among them to show how these entities are connected. The essential capabilities of NGSON architecture are service composition, service discovery and negotiation, service routing, context information management, content delivery, and service policy decision to enforce service and transport QoS to the underlying networks [230].

### 6.1.13. ONAP

ONAP [232] was formed as a merger of the open source version of AT&T's ECOMP and the Open-Orchestrator projects. ONAP provides a comprehensive platform for real-time, policy-driven orchestration and automation of physical and VNFs that will enable developers and service providers to automate and support complete life-cycle management of new services. It consists of both design-time (on-boarding new types of services) for VNF and PNF at run-time. The ONAP decouple the details of specific services and technologies from standard information models, core orchestration platform, and generic management engines (for discovery, provisioning, assurance, etc.). Furthermore, it marries the speed and style of a DevOps/NetOps approach with the formal models and processes an operator require to introduce new services and technologies [201]. It leverages cloud-native technologies including Kubernetes to manage and rapidly deploy the ONAP platform and related components. This is in contrast to traditional OSS/management software platform architectures where hardcoded services and technologies required lengthy software development and integration cycles to incorporate changes.

## 6.2. Global standardization efforts on 5G network slicing

This section provides the standardization efforts on network slicing from different telecom industry and bodies as shown in Fig. 21. The current discussions in the industry have been focused on the concept and requirements of network slicing, analysis of its impact on different levels or layers of the network stack (e.g., CN, the RAN,). For example, from the perspective of vertical industries, the 5G Automotive Association (5GAA)[21] is working with other companies from the automotive, technology, and telecommunications industries (ICT) to develop E2E solutions for future mobility and transportation services. To date, the first workstream was established in 5GAA WG5 to understand the business model aspects of network slicing in the automotive industry. Note that, manufacturing industry organizations like Zentralverband Elektrotechnik und Elektronikindustrie (ZVEI)[22] and the Industrial Internet Consortium (IIC)[23] have been actively engaging in the development of 5G-based smart manufacturing solutions.

---

[20] https://standards.ieee.org/standard/1903_2-2017.html.

From the operator's point of view, telecom industry organizations like the GSMA and NGMN have been working on exploring the concepts, business drivers and high-level requirements of E2E 5G network slicing. The GSMA Network Slicing Taskforce (NEST) project was initiated to harmonize slicing definition, identify slice types with distinct characteristics and consolidate parameter and functionality requirements [14]. The NGMN Alliance was among the first to introduce the concept of network slicing named "5G slicing" as stipulated in its white paper [75]. Since then the NGMN has been developing, consolidating and communicating 5G network slicing requirements and its architecture. The TM Forum ZOOM project[24] described in Section 6.1.11 has been working on developing business models and scenarios with regards to service providers, vertical industries, and other potential 5G network slicing consumers. Standard Developing Organisations (SDO) are not behind in the standardization of 5G network slicing. As of today, technical specifications of various domains has been defined by different SDOs, including the (a) Radio Access Network (RAN) and Core Network (CN) (e.g., in 3GPP), Transport Network (TN) (e.g., in BBF and IETF), (c) Application Layer (e.g., 3GPP). We provide the major highlights of these standardization activities in the next section.

### 6.2.1. ETSI

Activities with regards to 5G network slicing from the ETSI spans across different working groups. With the vision of enabling full automation in terms of deployment, configuration, assurance, delivery, and optimization of 5G network services, the ETSI Zero touch network and Service Management Industry Specification Group (ZSM ISG) has been actively working to resolve the 5G E2E network slicing management issues. The ETSI NFV ISG [233] provides technical solutions for network slicing resources such as computing and storage. The ETSI recognize SDN and NFV as enablers for multi-tenant and multi-domain environments in 5G infrastructure. To date, seven use cases have been defined including: single operator domain network slice, network slice instance creation network slice subnet instance creation, network slice instance creation, configuration and activation with VNFs, the priority of NSI for re-allocating the limited resources, network slice as a service and network slice instance across multiple operators.

### 6.2.2. 3GPP

3GPP is considered the forefront ambassador for 5G network slicing standardization activities. This is so because it consists of many working groups related to network slicing. The 3GPP SA1 defines use cases and requirements while 3GPP SA2 specifies the first system architecture choice to support network slicing. SA3 WG specifies security capabilities of E2E network slicing that require triggering, and coordination with the ETSI ISG NFV on the isolation of network slices [14]. The 3GPP SA5 defines the management of slices in coordination with other relevant SDOs to generate a complete E2E network slice. It is worth noting that, the 3GPP RAN1/2/3, is responsible for the RAN slicing awareness features [79].

### 6.2.3. ITU-T

The IMT2020 [234] is a proposal from ITUT-T that supports diverse service requirements with an E2E network slicing functionality to provide dedicated logical networks to customers. Specific functionality include: (a) network capability exposure, (b) softwarization everywhere leveraging existing tools such as SDN and NFV, (c) different mobility and diverse E2E QoS (data rate, reliability, latency etc.) requirements, (d) edge cloud support (MEC) with distributed content and services, and (e) separation of control plane (CP) and user plane (UP) functions, allowing independent scalability and evolution. Standardization activities from ITU-T SG13 [235] include the development of requirements and a framework for network management and orchestration with regard to vertical (service to network resources) and horizontal slicing. The ITU-T SG13 also defines an independent management of each plane (service, control data) and association of a user with multiple type of slices which is very closely coupled with 3GPP work. It further define high-level technical characteristics of network softwarization for IMT-2020, and data plane programmability (allow tenants of slices to provide top design tight integration data plane). ITU-T SG15 has developed an architecture of Slicing Packet Network (SPN) for 5G transport along with network slicing requirement for a transport network with SDN. A snapshot of this concept is illustrated in Fig. 22.

The proposed architecture can support all kinds of services in the metro network such as wireless backhaul, enterprise Ethernet-Line (point-to-point) and Ethernet LAN (multipoint-to-multipoint) services, and residential broadband. It can also support a simple switching mechanism to achieve low latency and low delay variation of E2E services.

### 6.2.4. ONF

The ONF [86,236] recognize that 5G will necessarily evolve in a brownfield while the SDN architecture will provide gradual migration and long-term coexistence with current management and signaling systems and network devices. The ONF was the first to apply SDN architecture to 5G network slicing as shown in Fig. 9. Since then, the functional requirements for Transport API (TAPI)[25] were specified. Potential use cases of TAPI include the integration of control and monitoring of optical transport network with higher level applications. This involves the support of network slicing enabling connectivity for high bandwidth or ultra-low latency 5G services through isolation and secure virtual subsets of the network [236].

### 6.2.5. BBF

The main activities of BBF are to define the slicing management architecture for TN and clarify the requirements for 5G bearer networks. In collaboration with the 3GPP, the BBF facilitates the transmission requirements from 3GPP and coordinate the interface requirements between the bearer slicing management system and 3GPP slicing management system [14]. This also includes recommending the technical definition for a specific interface and the

---

corresponding slicing creation and management processes. The BBF has been working on defining specifications of Fixed Access Network sharing (FAN). Some of the BBF standardization activities include broadband network infrastructure sharing among service providers to support management and control of resources. That way, operators can run their differentiated service operations at the minimum cost [164].

### 6.2.6. IETF

Standardization activities in the IETF involve the specification of general requirements and the development of 5G network slicing architecture, network slice management and orchestration mechanisms including lifecycle management to coordinate E2E and domain orchestration. It is important to mention that, some of the recent works in the IETF include: applicability of Abstraction and Control of Traffic Engineered Networks (ACTN) to network slicing, gateway function for network slicing[26], management of precision network slicing and packet network slicing using segment routing.[27]

### 6.3. Proof of concepts (PoC) for 5G network slicing

There are have been a significant number of PoC ranging from RAN slicing [237], multi-tenant hybrid slicing [238], 5G edge resource slicing [239] etc. Koutlia et al. [237] employ 5G-EmPOWER[28] to demonstrate a RAN slicing for multi-tenancy support in a WLAN. A hypervisor is proposed that can assign every AP the appropriate resources per tenant according to their traffic requirements. Pries et al. present a demo on 5G network slicing using an example of a health insurance provider who can request a network slice to offer services to customers. Liang et al. [241] propose a network slicing system for multi-vendor multi-standard Passive Optical Network(PON) in 5G networks while Guo et al. [238] demonstrate a multi-tenant scheme with unified management of various resources and resources isolation.

Zanzi et al. [242] proved the feasibility and reliability of Over-booking Network Slices (OVNES) solution in real cellular deployments. Costanzo et al. [243] present an SDN-based network slicing in C-RAN. Authors demonstrate a spectrum slicing prototype that shares the bandwidth resource efficiently among different slices while considering their requirements. Through a PoC, Boubendir et al. [239] illustrate on-demand creation and deployment of network slices dynamically over multiple domains for live content services in a stadium. A network operator can achieve the federation of access and edge resources owned by private third-party actors through B2B relationships. Capitani et al. [244] demonstrate the deployment of a 5G mobile network slice through the 5G-Transformer architecture experimentally. Raza et al. [245] present a PoC demonstration of an SDN/NFV-based orchestrator that enables resource sharing among different tenants. The profit of an infrastructure provider is maximized by the proposed orchestrator using a dynamic slicing approach based on big data analytics. From the industry, a slice-management function and network slices based on service requirements were demonstrated by DOCOMO and Ericsson in [246], enabling widely varying services to be delivered simultaneously via multiple logical networks. The PoC shows how 5G services could be connected flexibly between networks according to a set of policies in order to meet specific service requirements for latency, security or capacity.

### 6.4. Summary and lesson learned

We provide in Fig. 23 a summary of open source orchestrators, standardization efforts and PoC for 5G network slicing. It is evident that the network slice orchestrators have been developed in recent years from the industry and academia mainly to enable the orchestration and management of resources in future 5G networks. The area of concern has been to support the development and evolution of NFV components and SDN infrastructure across different levels of 5G ecosystems. Projects such as OPNFV focus on multi-vendor NFV components to ensure performance targets and interoperability by accelerating emerging NFV products and 5G services. Projects such as CloudNFV, Cloudify and T-NOVA leverage the benefits of NFV-based cloud management solutions and SDN to enable automated provisioning, monitoring, configuration, and efficient operations of NFaaS on top of virtualized 5G network infrastructures. Table 9 provides a summary and a comparison of different orchestrators for enabling network slicing on 5G networks. Based on the presented PoC in Section 6.3 and standardization efforts in Section 6.2 from the telecom industry and different bodies, it is evident that network slicing concept is an appealing solution for meeting vertical requirements and user's demands on 5G systems.

## 7. 5G network slicing orchestration and management

This section presents the management and orchestration approaches in 5G network slicing across different administrative domains while supporting multiple tenants. We first present the management and orchestration of network slices in a single domain followed by a comprehensive survey of approaches that consider management and orchestration of slices in multiple domains. The last part of this section covers the network slicing management and orchestration in edge and fog networks. According to the analysis of the industry and standardization resources [14,75,83,85,86,94,247,248], the requirements for an E2E management and orchestration in 5G network slicing include: flexibility, customization, simplification, exposure, elasticity, cloudification, legacy support, lifecycle management, automation, isolation and multi-domain and multi-tenant support. The identified requirements illustrate the need for centralized management and orchestration of network slice instances. This is so because the current management elements, network managers and OSS/BSS have no such capabilities. As a starting point for example, popular SDN controllers (e.g., [103,249,250]) are used for controlling the network resources within a single domain. Although, these solutions are very potential in managing network resources in a centralized manner, they do not provide standard virtualization features which allow, for example, to expose the PNFs like a physical OpenFlow-enabled switch as VNFs to the layers above.

### 7.1. Single domain management and orchestration

There exist a number of research works on network slicing management focusing on orchestrating resources from either a single type of network infrastructure resource domain (e.g., NFV) [256], a single network domain type (e.g., RAN) [257], or using a single type of resource domain manager (e.g., SDN controller) [176,258]. Iovanna et al. [259] provide a novel information model as part of the network abstraction that describes the flexible capability of the underlying transport network. Moreover, the proposed model defines a solution that provides efficient intra-domain resource management and allows a transport cost optimization using an E2E service routing algorithm. It provides dynamic and carrier-grade E2E transport connectivity combining heterogeneity,

---

**Table 9**
Summary of Orchestration Enabling technologies in 5G Networks [201] .

| Orchestrator | Technology | Organization | Objectives | Technology Features | Management Features |
|---|---|---|---|---|---|
| OpenMANO | SDN, NFV | Telefonica | To provide a practical implementation of the NFV MANO reference architecture | OpenMANO, OpenVIM, REST API | — |
| OSM | Cloud networks and services | ETSI NFV | MANO with SDN control, multi-site/multi-VIM capability | OpenMANO, OpenVIM, JuJu, OpenStack | — |
| OPNFV | NFV | Linux Foundation | To facilitate the development of mult- vendor NFV solutions across various open source ecosystems | OpenStack, OpeDaylight | — |
| ECOMP | SDN/NFV, Cloud and legacy networks | AT & T | Software centric network capabilities and automated E2E services. | TOSCA, YANG, OpenStack, REST-API | Improved OSS/BSS, service chain, policy management |
| T-NOVA | Network services and virtual resources | European Union | Network function as a service. | OpenStack, OpenDaylight | OSS/BSS, service lifecycle |
| OpenBaton [121] | Heterogeneous virtual infrastructures | FOCUS | Enables virtual network services on a modular architecture. | TOSCA, YANG, OpenStack, Zabbix | Event management and auto-scaling |
| Cloudify | NFV, Cloud | Gigaspaces | A multi-cloud solution for automating and deploy network services data centers. | TOSCA, OpenStack, Docker, Kubernetes | Service chaining, OSS/BSS |
| ZOOM | NFV and cloud services | TM Forum | Monitoring and optimization of Network Functions-as-a-Service (NFaaS). | — | Improved OSS/BSS |
| CloudNFV | SDN/NFV enabled cloud services | European Union | Enables the NFV deployment in a cloud environment | OpenStack, TM Forum SID | Service chaining and OSS/BSS |
| HP OpenNFV | NFV | European Union | An NFV-architecture that allocates resources from an appropriate pool based on global resource management policies. | Helion OpenStack | — |
| Intel ONP | SDN & NFV | Intel Corporation | Accelerates the adoption of SDN and NFV in telecom, enterprise, and cloud markets. | OpenStack, OpenDaylight | — |
| M-CORD | SDN, NFV- edge clouds for mobile networks | ON.Lab and partners | Anything as a Service, Micro-services architecture. | ONOS, OpenStack, XOS | Real-time resource management, monitoring/analytics, service chaining |
| OPEN-O | SDN, NFV and Cloud | Linux Foundation | Enable an E2E service agility across SDN, NFV, and legacy networks via vendor-specific data models (e.g., TOSCA and YANG) | TOSCA, YANG, OpenStack, REST-API, OpenDaylight, ONOS, Multi-VNFM/VIM | Improved OSS/BSS, service chain, policy management |
| ExperiaSphere | SDN, NFV & Cloud | CIMI Corporation | Flexible service model | USDL, TOSCA | Service events, derived operations |

NETCONF is the Network Configuration Protocol [216] that provides mechanisms to install, manage, and delete the configurations of network devices. YANG [217,218] is a data modeling language for configuration data, state data, remote procedure calls, and notifications for network management protocols, e.g., NETCONF and RESTCONF [219]. TOSCA [220] is a language that describes the relationships and dependencies between services and applications that reside on a cloud computing platform.

elasticity, and traffic engineering capabilities in each domain. Mohammed et al. in [260] present an SDN controller that performs orchestration of network connectivity resources. The controller takes appropriate actions whenever data delivery degradations (e.g., congestions) are detected in the network paths. That way, the proposed orchestrator provides elasticity at the level of data service delivery chain. Chatras et al. [261] propose to add a "Slice Controller", the functional block within the OSS/BSS responsible for interacting with the NFV management and orchestration system to control slicing. The Slice Controller is a consumer of the REST APIs exposed by the NFV which is responsible for assigning services to network slices, managing the life-cycle of these slices and mapping network slices to NFV network services. A SDN-based Technology Readiness Level (TRL-7) prototype for cross-domain slicing orchestration operations for the case of industrial applications with flexible QoS requirements is introduced in [262]. Talarico et al. [263] present an integrated service model that enables dynamic service discovery in the context of 5G network slicing using the concept of CloudCasting protocol [264]. The proposed approach provides several advantages such as enhanced scalability, accommodation of heterogeneous connectivity, lightweight signaling to establish service discovery and distribution, service isolation and prompt reaction to mobility. Celdrán et al. [263] propose a mobility-aware architecture that combines NFV/SDN techniques with innovative components responsible for managing and orchestrating the network slices. Kammoun án et al. [265] propose a new mechanism for the admission control for network slicing management in SDN/NFV environment. Authors introduce a network orchestrator that determines whether an existing slice can serve new users' requests demands or not. Wen et al. [266] investigate the robustness of network slicing mechanisms for the next generation of mobile networks. Authors propose an optimal joint slice recovery and reconfiguration algorithm for stochastic traffic demands by exploiting robust optimization where slice remapping is employed for re-selecting VNFs and links in order to accommodate the failed demands. Kotulski et al. [267] provide a constructive approach to E2E slice isolation in 5G networks. Kukliński et al. [268] propose an approach for a single domain consisting of management and orchestration of slices which are distributed into several functional blocks. The NFV MANO complaint orchestration mechanisms employed in [261] for enabling 5G network slicing are adapted without any modifications. A high-level overview of single domain management and orchestration is shown in Fig. 24. The first group of Global OSS/BSS building entities include the generic eTOM functions and portals for operators and tenants. The second group (Single Domain OSS) is responsible to provide a single domain slice management and orchestration. Generally speaking, the Global OSS/BSS is a logically centralised master block that drives the behaviour of the entire system including the MANO compliant orchestration. Users and operators' policies are analysed by the Slice Configurator as soon as the slice request is made. The NFVO Support block is responsible to create the Network Slice Description (NSD) that is used by the NFVO for slice deployment. It also keeps the catalogue of network slices. The Domain Manager performs resource allocation to slices based on their demands and priorities. As shown in Fig. 24, the Slice Manager (SM) is an entity that handles faults and performance of a slice or sub-slice. In co-operation with the Global OSS/BSS, which plays the master role in the overall management and orchestration, the SM is also responsible for managing the sliced network. It is important to mention that, all requests regarding slice creation and termination as well as access to current and historical data related to a particular slice from tenants are made through the Global OSS Tenants Portal. One of the shortcomings in single domain management and orchestration in 5G network slicing is a scalability issues [13]. This is so because, only a single network domain type (e.g., RAN) [257], or using a single type of resource domain manager is used. Scalability problems can be solved by using multi-domain orchestrator (MdO) implementation. The MdO represents the first step for a fast and automated network slices provisioning over multiple-technologies spanning across multiple-operators [158,255].

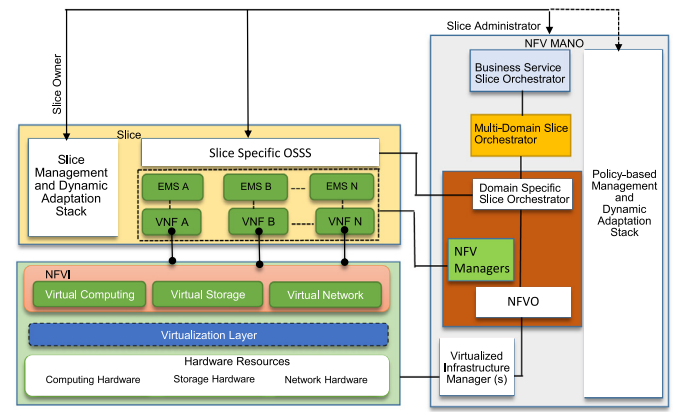

**Fig. 18.** Illustration of 5G!Pagoda network slice orchestration and management architecture.

### 7.2. Multi-domain orchestration and management

Multi-domain provides a realization of E2E management and orchestration of resources in 5G sliced networks [269]. The implementation of multi-domain in 5G networks is set to enable the interaction of multiple administrative domains at different levels with different service and infrastructure providers. It ensures that service requests from different domains are mapped into multi-operator and multi-technology domains while matching each service ELA requirements [255]. Perez-Caparros et al. [270] was among the first to design the MdO use cases and its requirements. This was followed by several research works [271–273] that suggested where and how to place VNF in multi-domain architectures. An initial analysis of multi-domain orchestration frameworks is given in [208]. Guerzoni et al. [255] present an E2E management and orchestration functional architecture for multi-domain 5G environments. The first implementation of the 5GEx MdO prototype obtained following an extension of this architecture is available in [158] where authors demonstrate how it is possible to create and deploy network slices in the context of a Slice as a Service (SlaaS) use-case based on a multi-operator scenario. Vaishnavi et al. [274] provide an experimental implementation of multidomain orchestration where multi-operator services can be deployed and monitor the service for ELA/SLA compliance over 5G networks. Dräxler et al. [275] propose a 5G Operating System (5GOS) that can provide control and management for services running on top of a multi-domain 5G infrastructure. In 5GOS, the control and manipulation of resources in different administrative and technological domains is done by domains specific SDN controller and NFV MANO systems. It is important to mention that, multi-domain orchestrators handle the life-cycle management of E2E slices across multiple administrative domains while domain-specific orchestrators build slices of the network, compute, and storage resources. As the vision of 5G!PAGODA, Afolabi et al. [254] propose a 5G network slicing architecture whereby slices of virtual mobile networks are created on-demand and customized according to the changing needs of mobile services using physical resources across multiple domains. All network slices in the 5G!PAGODA showed in Fig. 18 (see Section 5.1.10) can be implemented following the slice template as illustrated in Fig. 25.

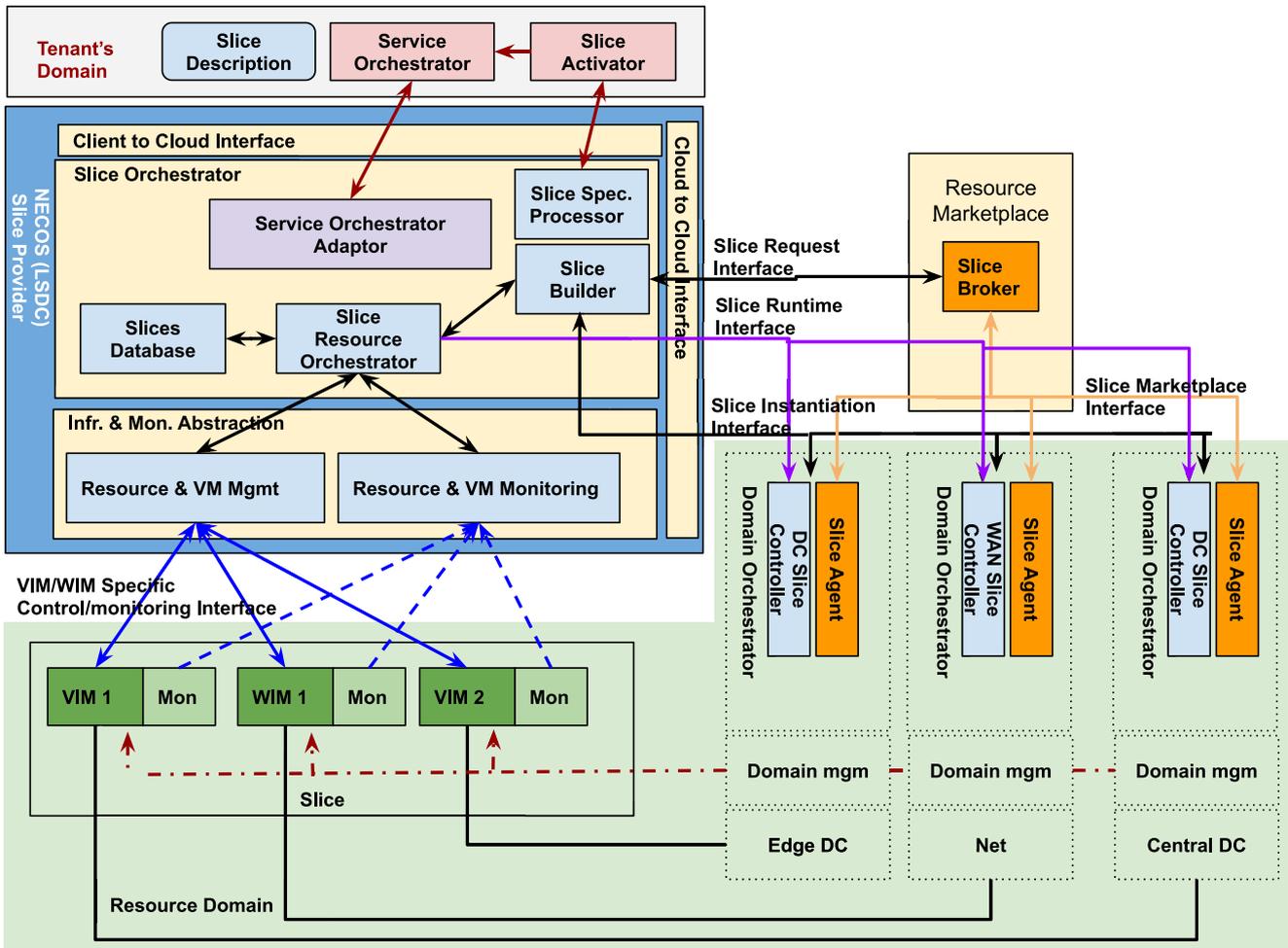

Fig. 19. The NECOS slice provider and data center / network provider role.

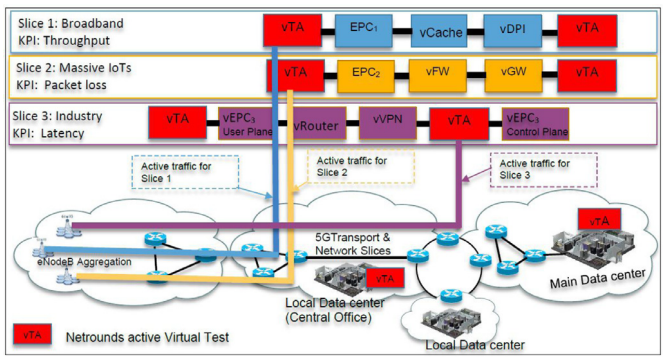

Fig. 20. Three network slices with different critical key performance indicators. The Massive IoT slice may include VNFs like vEPC, vFW, and vGW. Packet loss is the KPI to meet SLA requirements for this slice. The industry slice may include VNFs such as vRouter and vVPN. Latency is the KPI for this slice to meet SLA requirements [210].

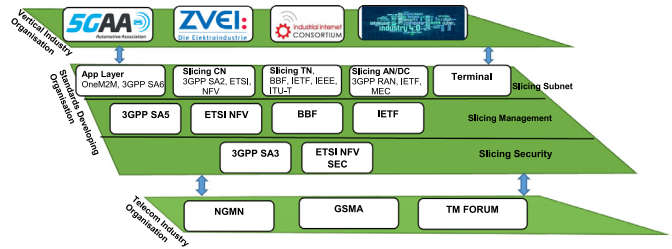

Fig. 21. Network slicing relevant industry groups and SDOs landscape.

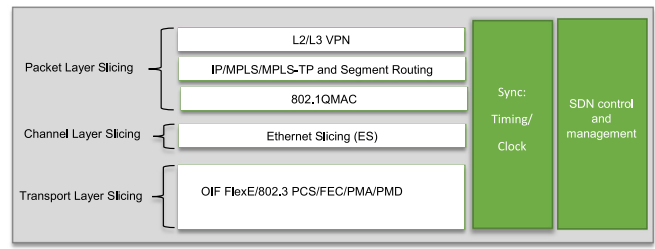

Fig. 22. High level overview of an architecture of Slicing Packet Network (SPN) for 5G transport.

In 5G!PAGODA design, a network slice consists of all the network components such as a RAN, transport, core network and different application enablers (e.g., video streaming optimizer). In order to optimize the network slice functionality, the RAN, for example, can be shared between multiple-domains and provide specific services to the end users in an efficient way using the lifecycle management plane [24]. It is worth noting that, different 5G slices are instantiated and run in isolation on top of the same infrastructure which can be operated and managed by multiple operators and providers (e.g., telecom operators, MVNOs, cloud providers, etc.). An essential module in 5G!PAGODA proposal is the

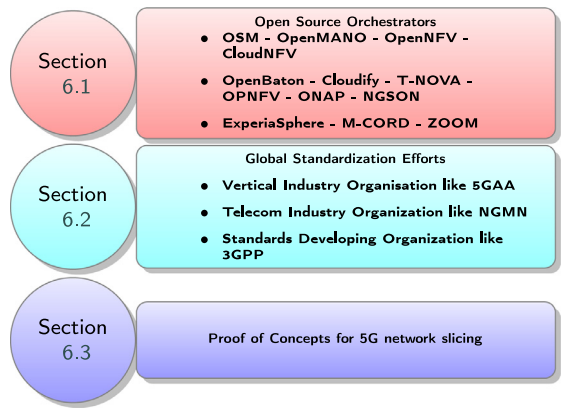

**Fig. 23.** A summary of Section 6.

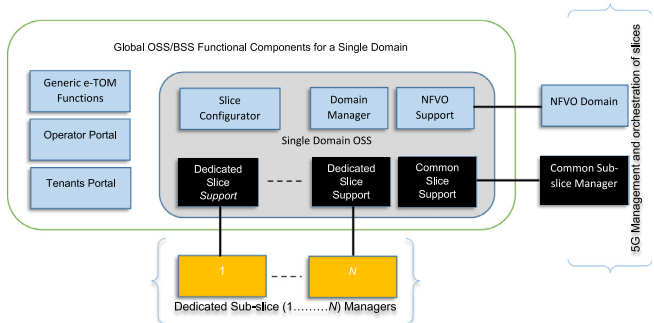

**Fig. 24.** High-level overview of management and orchestration in a single domain. The "Enhanced Telecom Operations Map (eTOM) is a reference framework that categorizes the business processes that a service provider will use.

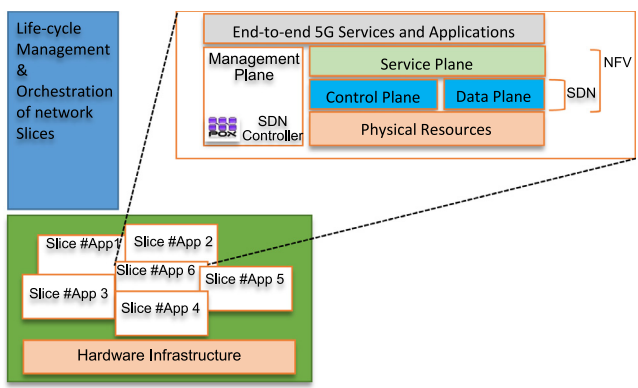

**Fig. 25.** 5G!Pagoda slice template and instantiated slices.

Resource Orchestration (RO) which have a global view of all the resources inside an administrative domain. This makes it easy to place VNFs and create related Network Function Forwarding Graph (NFFG) across different resources (e.g., across multiple data centers) efficiently [254].

The concepts of vertical and horizontal slicing[29] is invoked by Li et al. [276] to enable 5G system with E2E network slicing. Authors provide a system architecture with E2E vertical and horizontal slicing followed by a discussion of promising technologies in

---

the air interface, the RAN and CN. In order to enable slicing in the RAN, authors suggest for each slice to have its RAN architecture where the control-plane (C-plane) and user-plane (U-plane) configuration can be tailored considering the slice-specific operation. That way, RAN operations (e.g., access control, mobile association and load balancing) schemes have to be slice-specific. This is different from the traditional operations in mobile networks where a cell-specific is considered [276]. NFV and SDN are technical enablers of network slicing in the CN such that, each CN slices is defined to support different services/applications. While that is the case, slice pairing functions are also defined to pair the radio, RAN, and CN slices with the endpoint of forming E2E slices. Katsalis et al. [277] propose 5G network slicing reference architecture for the problem of multi-domain NFV orchestration with a specific focus on LTE networks. In [93], authors present an SDN/NFV-based control plane that enables multi-tenancy through network slicing. The architecture provides an efficient allocation of transport network resources to multiple tenants [196]. On multi-operators network sharing perspective, Caballero et al. [278] addressed RAN slicing of multiple tenants managed by multiple virtual wireless operators and service providers. Based on a weighted proportionally fair objective, authors consider dynamic resource allocation to achieve desirable fairness across the network slices of different tenants and their associated users. DASMO [279] is multiple in-slice autonomous management platform that enables the creation of distributed and automated network slicing. DASMO consists of an embedded management intelligence of slice nodes and allows for local (e.g., at the slice level) management decisions. That way, DASMO reduce delays and provide efficient management of network traffic. Raza et al. [280] present a comprehensive assessment of problems given by dynamic slicing in a 5G transport network. The results based on a mixed integer linear programming (MILP) formulations and heuristic algorithms indicate that both re-sizing and re-mapping of slices provide efficient utilization of physical network resources. As the vision of SliceNet presented in Section 5.1.3, Wang et al. [281] demonstrate a QoE-driven 5G network slicing framework focusing on cognitive network management and control for E2E slicing operation and slice-based/enabled services across multiple operator domains. Authors in [282] investigate the resource allocation problem of achieving maximum capacity with the transmit power, allocated bandwidth as part of the constraints in a sliced multi-tenant network. A network slice manager in SONATA Service Platform (SP) [256] is proposed in [282] for multi-site NFVI-PoP that supports multi-tenancy. NESMO is among the recent network slicing management and orchestration architecture proposed in [101] that extend the 3GPP management reference framework [83]. NESMO consists of the *Network Slice Design* and *Multi-Domain Orchestrator* components that are needed to design, deploy, configure and activate an NSLI in multiple network infrastructure resource domains. It also consists of a *multiple network infrastructure resource domains* that can manage not only NFVI but also different types of infrastructure resources [101]. Taleb et al. [253] propose a multi-domain management and orchestration architecture for 5G network slicing that can provide services across federated domains. The proposed architecture for multi-domain 5G network slicing shown in Fig. 26 consists of the following functional entities: Multi-domain Service Conductor Stratum, Domain-specific Fully-Fledged Orchestration Stratum, Sub-Domain Management and Orchestration (MANO) and Connectivity Stratum, and Logical Multi-domain Slice Instance stratum. We provide a brief description of each entity herein.

- *The multi-domain service conductor stratum*: Is responsible for mapping all service requirements of different multi-domain requests to their respective administrative domains. It consists of two modules namely, *Service Conductor (SC)* and *Cross-domain*

---

[29] Vertical slicing (VS): is a viable solution that provides different needs of vertical industry and markets with resource sharing mechanisms among services and applications. Horizontal slicing (HS) - extends the capabilities of the mobile device by providing resource sharing mechanisms among network nodes and devices [276].

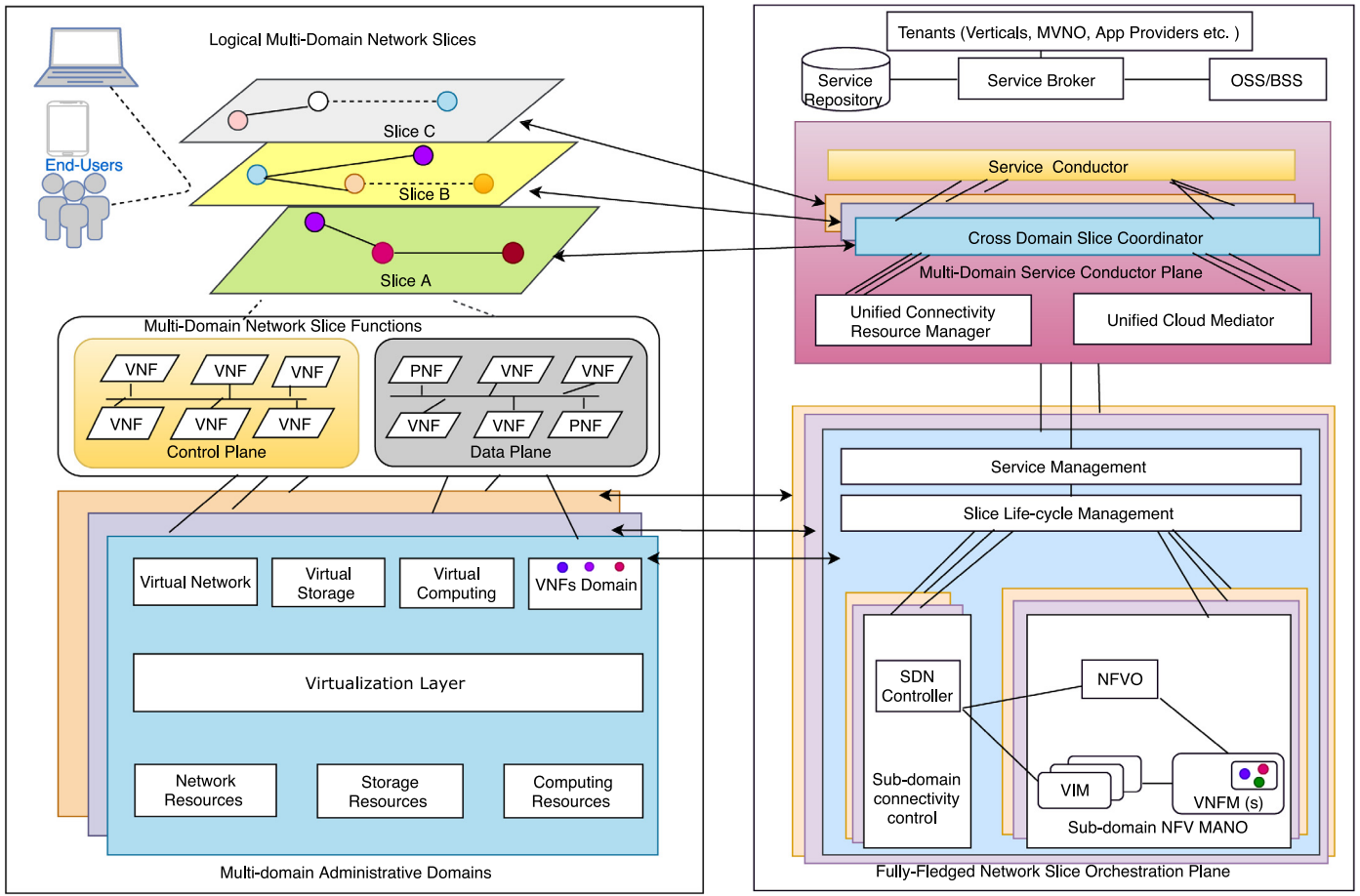

**Fig. 26.** A multi-domain slicing architecture in 5G networks (adapted from [253]).

*Slice Coordinator (CSC).* The SC performs re-adjustment of network resource in different federated administrative domains during network performance degradation or service policy provisioning changes.

- *Domain-specific fully-fledged orchestration stratum*: It allocates internal domain resources for establishing a federated NSI and provides the corresponding LCM using the *service management, Sub-domain NFV MANO, Sub-domain SDN Controller* and *slice life-cycle management* functions. The former functional entity analyses the slice request received from the Cross-domain Slice Coordinator and identifies the RAN and core network functions, including value-added services. The latter is used to identify the appropriate network slice template from an associated catalog and forms a logical network graph which is then mapped to the underlying resources (e.g., compute, storage and network) corresponding to a technology-specific slate [253]. Sub-domain SDN controller provides the network connectivity and service chaining among the allocated VNFs that connects remote cloud environments using PNFs.

- *Service Broker Stratum (SBS)*: This is introduced as a service broker in the functional plane to handle all incoming service request from application providers, MVNO and different verticals. The SBS is responsible for the management of NSI revenue that involves charging and billing of slice owners. It also performs network slicing admission control and negotiation by considering different service requests from various administrative domains.

- *Sub-domain Infrastructure Stratum (SIS)* : It consists of the physical and virtual infrastructure (e.g., VNFs, virtual resources, virtualization layer, and physical infrastructure):

Taleb et al. [253] also provide a discussion on multi-domain network slice orchestration and management procedures for multi-domain network slice configuration and multi-domain network slice modification. Three fundamental design challenges relevant to the realization of service management in 5G network slicing include: resource isolation and sharing, service management interfaces & service profiling and service-based management plane. For service management interface, the RESTfull models like L3SM/L2SM [283,284] and NFVIFA Os-Ma-Nfvo[30] are currently being considered for facilitating the information exchange. These models are used for programmability purposes to provide control capabilities among different administrative and technology domains and third parties such as verticals. However, the problem with these models is the lack of resiliency and performance measurement capabilities, multi-domain connectivity and control considerations on federated resources. It is therefore argued that new data models have to be developed that can analyze and map service requirements of the corresponding slice into the relevant cloud and networking resources. Besides that, service profiling algorithms for optimizing the mapping of allocated resources are highly needed in future 5G network slicing environments [253]. Authors in [285] propose an E2E slices architecture platform that exploits feedback information from mobile network slices to make orchestration decisions via a hierarchical control plane. Efficient management mechanisms are proposed that per-form resource reservation and slice admission control decisions across all mobile network domains [285].

---

[30] 5ETSI GS NFV-IFA, Os-Ma-Nfvo reference point, Interface, and Information Model Specification, Oct. 2016

### 7.3. Network slicing management in edge/cloud and fog computing

MEC promises to offer an environment characterized by high bandwidth and low latency for applications and content providers. An efficient QoE monitoring and management approach that is aware of the RAN type, cell topology and resource allocation for adapting to the service delivery characteristics and end user's QoE demand is proposed in [62]. A joint heterogeneous statistical QoS/QoE provisioning for edge-computing based Wi-Fi off-loading over 5G mobile wireless networks is presented in [286]. Ge et al. [287] propose an Edge-based Transient Holding of Live Segment (ETHLE) strategy to achieve seamless 4K live streaming experience by eliminating buffering and substantially reducing initial startup delay and live stream latency. Husain et al. [288] propose a MEC with network resource slicing for IoT devices.

Truong et al. [289] is among the earliest to propose an SDN-based architecture that supports Fog Computing in the context of Vehicular Adhoc Networks (VANETs). Authors demonstrate the benefits of their proposed architecture using two use-cases in data streaming and lane-change assistance services. SDN controller is used to control and manage resources/services and to optimize their migration and replication. Bruschi et al. [290] introduced a network slicing scheme that supports multi-domain fog/cloud services. The experimental results show that the number of unicast forwarding rules installed in the overlay network significantly drops compared to fully-meshed and OpenStack cases. Based on the NFV MANO framework and OpenFog Consortium (OFC)[31] reference architecture, Lingen et al. [291] introduce a model-driven and service-centric architecture that addresses technical challenges of integrating NFV, Fog, and MEC on 5G networks. One of the use cases used in their pilot study is the physical security of Fog nodes using a two-layer abstraction model along with IoT-specific modules. In [292], an ONOS SDN controller is used to design a Fog operating system architecture called FogOS for IoT services while Diro et al. [293] propose a mixed SDN and Fog architecture that gives priority to critical network flows while taking into account fairness among other flows in the Fog-to-Things [294] communication. The proposed Fog architecture can satisfy QoS requirements of heterogeneous IoT applications.

Recent development with regards to 5G network slicing in MEC and/or cloud/Fog computing domain include [295–299]. Authors in [295] present a MEC-enabled 5G architecture that supports the flexible placement/migration of network and application VNFs. NFVO orchestrates the VNFs with admission control and management capabilities that can manage the NFs and resources on-the-fly. While streaming live content services in a stadium, Boubendir et al. [296] illustrate on-demand creation and deployment of network slices dynamically over multiple administrative domains. Zanzi et al. [297] introduce the concept of MEC broker (M$^2$EC) that leverage the network slicing paradigm to allow renting part of MEC facilities. This in turn, enables both the system provider and the MEC tenants to expand their business opportunities. M$^2$EC is an entity that exposes administration and management capabilities while handling heterogeneous tenant privileges. That way, it optimally allocates requested resources in compliance with the tenants SLAs/ELAs. Amemiya et al. [299] propose a novel slicing method for softwarized BS to isolate a low latency slice from a broadband slice. The Authors' evaluation indicates reasonable resource isolation and minimal latency in the proposed method. By adapting the NFV to MEC network, authors in [298] present an SFC slicing scheme that utilizes the popularity of 5G network services to decide the number of replicas (of network services) which can minimize service time. The presented works for 5G network slicing using Fog/Edge and cloud computing indicates that, a lot of works have to be done with regards to the development of algorithms for QoS/QoE monitoring and management of resources in softwarized infrastructures.

### 7.4. RAN slicing for multi-Service 5G networks

RAN slicing is stemmed from the RAN sharing concept such as Multi-Operator RAN (MORAN) and Multi-Operator CN (MOCN). The MORAN approach enables sharing the same RAN infrastructure but with dedicated frequency bands for different operators. The MOCN concept enables sharing the spectrum among operators as standardized by 3GPP [300]. These solutions can utilize the available radio resources efficiently and are widely surveyed as network virtualization substrate (NVS) [301,302]. Radio resources for different resource provisioning approaches can be virtualized in order to coexist several mobile MVNOs in a single physical RAN. Mahindra et al. [303] propose a NetShare approach as an extension to the NVS solution. A central gateway-level component is applied to ensure optimization and isolation of resources distribution for each entity. The CellSlice architecture is proposed by Kokku et al. [304] as a gateway-level solution for slice-specific resource virtualization that can indirectly impact individual BS scheduling decision. The application-oriented framework called AppRAN is presented in [305] that defines a serial of general applications with distinct QoS guarantees. Aijaz [306] proposes a Hap-SliceR radio resource slicing framework which is based on the reinforcement learning approach that considers slice utility requirements and resource utilization. However, this approach is only focused on resource customization for haptic communication. It is worth noting that, RAN virtualization [307–309] generally provides functional isolation in terms of customized and dedicated control plane functionalities for each MVNO. The above approaches consider either functional isolation or radio resource sharing while less research attention is given to satisfying the concerns of functional isolation and radio resource sharing simultaneously.

Several 5G RAN design requirements, and paradigms have to be fulfilled as stipulated by Marsch et al. [310] in order to enable the RAN slicing concept. For example, based on the underlying principles of SDN/NFV, cloud computing, and software engineering, future RAN design patterns are presented in [257]. Moreover, the RAN 3GPP explores slicing realization principles including RAN awareness slicing, QoS support, resource isolation, SLA enforcement among others [311,312]. However, the software-defined RAN (SD-RAN) concept that decouples CP processing from the UP processing can enable RAN slicing principles. In line with SD-RAN concept, several works argue the level of centralization of CP functionalities. This include the fully centralized architecture proposals such as OpenRAN [313] and SoftAir [199] that may face the challenge of real-time control given the inherent delay between the controller and underlying RAN. In order to perform the control functionalities through the APIs, the SoftMobile [314] abstracts the CP processing in various layers. The control functions are refactored statically into the centralized and distributed functions in the SoftRAN [315] architecture based on the central view requirement and time criticality. The OpenRadio [316] and PRAN [317] are pioneered to decompose the overall processing into several functionalities that can be chained for the UP programmability and modularity. An SD-RAN platform is realized by FlexRAN [318] to implement a custom RAN southbound API through which programmable control layer can be enforced with different levels of centralization, either by RAN agent or the controller.

---

[31] The OFC offers uniform management of IoT services that span through the cloud to the edge network.

The RAN slicing works are initiated based on the enablers mentioned above. The RadioVisor [130] can isolate the control channel messages and primary resources (e.g., CPU and radio resource) to provide the customized service for each slice. FLARE [319] is a full isolation solution with different virtual base stations (vBSs) that represent different slices. The only drawback of this approach is the lack of multiplexing benefits in the radio resource allocation since the spectrum is partitioned disjointly. Again, this work does not consider multiplexing and network function sharing. Peter et al. [13] propose to separate the radio resource scheduling of a BS into the intra-slice scheduler and inter-slice scheduler. However, only a portion of functions are isolated, and the resource abstraction/virtualization is not included in this work. Using a novel resource visor for each slice, authors in [320] proposes a RAN slicing architecture that allows Radio Resource Management (RRM) policies to be enforced at the level of Physical Resource Blocks (PRBs) through providing the Virtualized Resource Blocks (vRBs). Nevertheless, neither resource customization/abstraction per slice request nor function isolation is considered in this work. In [321], different approaches to split radio resources are compared in terms of the resource granularity and the degrees of isolation and customization. The resource multiplexing capability among slices is not considered. The concept of BS hypervisor is introduced by Foukas et al. [322] to share the radio resources and isolate slice-specific control logics simultaneously. This work exploits the prerequisites of resource virtualization and function isolation, but customization and multiplexing of CP/UP functions in both disaggregated and monolithic RAN deployments are not considered. Physical resource partitioning based on the service descriptions to flexibly share RAN functions over different network layers is achieved in the proposed RAN slicing framework [323] – however, no any resource virtualization and multiplexing considered in this work.

### 7.5. Summary and lessons learned

This section has presented the management and orchestration approaches in 5G network slicing for single and multiple administrative domains. The network slicing management in edge/cloud and fog computing as well as RAN slicing for multi-service 5G networks is given. Table 10 summarizes the details of the quantitative analysis, showing the functional and operational features of some approaches discussed above from major 5PPP research projects (5G-EX, 5G!PAGODA, and 5G-NORMA) and those from standard bodies, namely the 3GPP, ONF-SDN, ITU-T, and ETSI NFV-MANO. We summarize in Table 11 the state-of-the-art solutions and a comparison of 5G RAN slicing approaches in terms of radio resource allocation model, control plane function, and user plane function. It is interesting to see how the academia and industry are pushing forward the implementation and adoption of 5G network slicing in different aspects. The common goal is to make sure that customers do not need individual agreements with different service providers or mobile operators for global service experience.

## 8. Future challenges and research directions

It is indomitable that the maturity and the inherent potentials of SDN, MEC, Fog/Cloud computing, and NFV are paving the way to transform the future 5G network infrastructure. Although, the network softwarization and network slicing concepts using SDN and NFV in 5G come with benefits (e.g., flexibility, agility, etc.) many challenges need to be resolved before the realization of this novel paradigm [328]. This section provides essential challenges and future research directions that need to be comprehensively resolved by the research community focusing on 5G network slicing.

**Table 10**
Comparison of Network Slicing Orchestration Architectures and their Offered Support.

| NS Architectures | Multi Tech. Domains | Unif. Cloud Med. | Unif. Connect. Mgmt. | Recursive Virtualization | Programmability | 3rdParty Contl./Orch | Federated LCM | Service Mang. | Service Chain & SDN | Broker; AC/Neg. | RAN Orch. | Multi-Tech.Domains |
|---|---|---|---|---|---|---|---|---|---|---|---|---|
| 3GPP TS 28.530 [251] | ✗ | ✗ | ✓ | ✗ | ✗ | ✗ | ✓ | ✓ | ✗ | ✗ | ✓ | ✓ |
| ITU-T Y.3011[77] | ✓ | ✗ | ✓ | ✗ | ✓ | ✓ | ✓ | ✓ | ✓ | ✓ | ✓ | ✗ |
| ONF TR-526 [86] | ✓ | ✗ | ✓ | ✓ | ✓ | ✓ | ✓ | ✗ | ✗ | ✗ | ✗ | ✓ |
| ITU-T Y.3011 [233,252] | ✓ | ✗ | ✗ | ✓ | ✓ | ✓ | ✓ | ✗ | ✓ | ✓ | ✓ | ✓ |
| Taleb et al. [233,253] | ✗ | ✓ | ✓ | ✓ | ✗ | ✓ | ✓ | ✓ | ✓ | ✗ | ✓ | ✓ |
| 5G-NORMA [174,176] | ✗ | ✓ | ✓ | ✗ | ✗ | ✓ | ✗ | ✗ | ✗ | ✓ | ✓ | ✓ |
| 5G!Pagoda [24,254] | ✗ | ✓ | ✗ | ✗ | ✓ | ✓ | ✓ | ✗ | ✗ | ✗ | ✓ | ✓ |
| 5G-Exchange[157,158,255] | ✓ | ✗ | ✓ | ✗ | ✓ | ✗ | ✓ | ✗ | ✓ | ✓ | ✗ | ✓ |

NS = Network Slicing; LCM = Life-Cycle Management; AC = Admission Control

**Table 11**
A Summary and comparison of RAN slicing Approaches [327].

| Reference | CP Function | UP Function | Radio Resource | Solution Level |
|---|---|---|---|---|
| Nakao et al. [319] | Dedicated | Dedicated | Dedicated Spectrum allocation | BS level |
| Ksentini and Nikaein [324] | Dedicated | Shared | Flexible between dedication and sharing | BS level |
| Foukas et al. [322] | Split into cell and user specific | Dedicated till PHY layer | Virtualized resource sharing | BS level |
| Foukas et al. [318] | Shared | Shared | Physical or virtualized resource sharing | BS level |
| Nikaein et al. [325] | Dedicated | Dedicated | — | Network-wide level |
| Kokku et al. [301] | — | — | Physical or virtualized resource sharing | BS level |
| Mahindra et al. [303] | — | — | Physical or virtualized resource sharing | Gateway and BS level |
| Kokku et al. [304] | — | — | Virtualized resource sharing | Gateway level |
| He and Song [305] | — | — | App-oriented Virtualized resource sharing | Gateway level |
| Aijaz et al. [306] | — | — | Learning-based virtualized resource sharing | Gateway level |
| Zaki et al. [308] | Dedicated | Dedicated | Physical resource sharing | BS level |
| Gudipati et al. [130] | Dedicated | Dedicated till programmable radio | Physical 3D resource sharing | BS level |
| Rost et al. [326] | Split into cell and user specific | Dedicated till real-time RLC | Physical resource sharing | BS level |
| Ferrús et al. [323] | Dedicated | Dedicated or shared till PHY | Physical resource sharing | BS level |

## 8.1. Network sharing and slicing in 5G

Moving from hardware-based platforms to software-based platforms could potentially simplify the multi-tenancy support where multiple services/applications from different vertical-specific use cases can be accommodated over a common SDN/NFV-based infrastructure in 5G systems as discussed in Section 7. Besides, evolving the network sharing paradigm to the concept of network slicing that enables multiple VNFs to be configured on the same NFV platform creates many management problems of large slices. It is worth mentioning that, although the dynamic resource sharing among slice tenants would make network resource utilization more efficient, it calls for intelligent scheduling algorithms that will allocate resources among these slices. Besides, the problems concerning NFs placement within the slice, intra-slice management, and inter-slice management still need significant efforts in order to achieve and realize the effectiveness of the network slicing concept in 5G networks.

Also, the problems related to the placement of network functions within a slice, slicing orchestration, or inter-domain services slicing also need to be further studied to achieve the effectiveness of network slicing. Again, another research direction that needs extensive explorations is related to isolation between slices, mobility management, dynamic slice creation, and security [329]. Concerning isolation, a set of consistent policies and appropriate mechanisms have to be clearly defined at each 5G virtualization layer. Moreover, in terms of performanceincluding QoS/QoE requirements have to be met on each slice, regardless of the network congestion and performance levels of other slices. Furthermore, in order to provide "Network as a Service" to the 3rd parties standardize interfaces for the information flow, requirements and management are needed.

## 8.2. End-to-end slice orchestration and management

Shifting from hardware-centric to software-centric paradigms using SDN and NFV in 5G networks will need changes on how networks are deployed, operated and managed. It also demands new ways on how resources are orchestrated while making sure that network functions are instantiated dynamically on-demand basis [39]. The ETSI MANO framework has already shown direction, with anticipated capabilities of life-cycle management and configuration of VNFs. Following that trend, other efforts that provide solutions for a management platform for VNFs, for example, the AT&T's ECOMP project [330], the OSM project [331], and ONAP project [232] implement the SO on top of NFVO. With ONAP, operators can synchronously orchestrate both physical and virtual

NFs. The OPNFV [222] creates a reference NFV platform to accelerate the transformation of enterprise and service provider networks. Related MANO frameworks and 5G architectures for 5G network slicing that considers the management and orchestration of both virtualized and non-virtualized functions are comprehensively elaborated in Sections 6 and 7. Despite these efforts, a significant challenge for 5G network slicing realization in terms of infrastructure and NFs as stipulated in [19] is how to move from a high-level description of the service to the concrete network slice. This calls for the development of domain-specific service/ resource description slicing languages that would allow the expression of KPIs, requirements, and characteristics of 5G network service KPIs. Flexibility and extensibility should be an essential ontologies for such network service/resource description [332] languages to support multi-vendor environments and accommodate new 5G network elements (e.g., new RATs).

NFV MANO frameworks such as OSM promise to realize the E2E 5G network slice. However, one of the major concerns is the holistic orchestration and management of different slices such that, each slice meets its service and ELAs/SLAs requirements while utilizing the underlying resources efficiently. This calls for sophisticated E2E orchestration and management plane and adaptive solutions that manage resources holistically and efficiently by making decisions (e.g., for slice generation and resource allocations) based on the current state of slices as well as their predicted future system state and user's demands [199].

## 8.3. Security and privacy challenges in 5G network slicing

The notion of sharing resources among slices may create security problems in 5G network slicing. This is so because network slices that serve different types of services for different verticals may have different levels of security and privacy policy requirements. This calls for the new development of 5G network slicing security and privacy protocols that consider the impact on other slices and the entire network systems while allocating resources to a particular slice(s). Also, security issues become even more complicated when 5G network slicing is implemented in multidomain infrastructures. In order to address this problem, security policy and efficient coordination mechanisms among different administrative domains infrastructure in 5G systems must be designed and developed. Generally speaking, efficient mechanisms have to be developed to ensure that any attacks or faults occurring in one slice must not have an impact on other slice [12,333]. That way, network sharing and slicing in 5G networks using SDN and NFV can be realized in the practical implementation without any security concerns.

### 8.4. Challenges of RAN virtualization in 5G network

RAN virtualization will be an integral part of 5G such that commodity IT platforms will have the potential to host cloud RAN networks [326]. RAN Slicing in virtualized 5G systems is still in its infancy stage. Applying containers such as Docker and VM-based solutions for RAN virtualization does not adequately address the problem. This is so because these solutions do not add any dimension of radio resources (e.g., spectrum or hardware) virtualization and isolation. While multiple RATs including emerging technologies like 5G new radio and NB-IoT) are expected to be a universal norm in 5G networks, it is of great importance for RAN virtualization approaches to be able to support multiple RATs. This calls for RAN slicing strategies that can flexibly support various slice requirements such as isolation and elastically improve multiplexing benefits (e.g., sharing) in terms of (a) network service composition and customization for modularized RAN, (b) flexibility and adaptability to different RAN deployment scenarios ranging from monolithic to disaggregated, and (c) the new set of radio resource abstractions. Again from the RAN virtualization perspective, the RaaS realization is another significant challenge in 5G network slicing paradigm. It is vital for RaaS paradigm in 5G networks to go beyond physical infrastructure and radio resource sharing. In that aspect, RaaS realization should have the capability to create virtual RAN instances on-the-fly with a simple set of virtualized control functions such as mobility management and scheduling of radio resources. With this in mind, the aim of RaaS realization in 5G network slicing should simultaneously suit individual slice/service requirements and ensure isolation between different slices (virtual RAN instances) [318].

### 8.5. Business model development and economic challenges for 5G network slicing

For many years, the telecoms industry has been providing services with guaranteed QoS to different customers using various levels such as the Integrated services (IntServ) [334] and the Differentiated services (DiffServ) models [335]. 5G networks will support new services such as ultra-reliable and low-latency communications (URLLC), enhanced mobile broadband (eMBB) and massive machine-type communications (mMTC). That way, 5G networks is positioned to meet the requirements of different vertical industrial applications and services in terms of latency and bandwidth. Industry verticals in 5G networks will be able to order a network slice through a North Bound Interface (NBI) of the MNO. To deploy and manage network slices, some of these verticals will rely on MNO. The charging/billing also for 5G verticals related to QoS, resources assigned to them and the overall performance will be different. 5G network slicing should provide solutions and drive new business models for delivering heterogeneous 5G- oriented services of interest for different industry verticals [336]. In this regard, multiple business models have to be developed for managing different services and applications as well as customers of the verticals in 5G networks.

There are three different possible business models for network slice commercialization, namely (a) Business to Business (B2B), (b) Business to Consumers (B2C), and Business to business to consumer (B2B2C) [57]. In the B2B domain, customized 5G network resources will be sold to enterprises by MNO. The full control of consumers in this service delivery chain is released to enterprises. The B2C domain involves customers purchasing customized 5G network resources based on their requirements. However customers does not take into considerations which MNO provides the requested resources. The B2C domain will allows quick monetization by evolving from video streaming towards making future personal lives digital. However, B2C poses a lot of challenges to MNO because the business model will need to change in terms of service model and and charging/billing metrics. B2B2C domain, online, or e-commerce, businesses and portals reach new 5G markets and customers by partnering with consumer-oriented product and service businesses. That way, MNOs have to provide customized network resources to a broker. The broker then engages with customers directly and gets more control of 5G network. The Network Slice Broker (NSB) in 5G systems proposed by Samdanis et al. [85] which can enable industry vertical market players, OTT providers and MVNO to request and lease resources from InPs dynamically via signaling means can be a starting point to realize the B2B2C model development.

Profit-maximization strategies of resource management for multi-tenant slices and the economics of 5G network slicing has been explored in recent studies [337]. Bega et al. [338] propose to optimize the 5G infrastructure markets in order to maximize the overall revenue of network slices. Resource allocation strategies for solving the profit maximization problem of a set of independent MVNOs that request slices from an MNO are analyzed in [339–341]. Authors in [342,342] propose an optimal resource allocation approaches within the context of 5G network slicing for enabling customization in multi-tenant networks in geographically distributed resources. Guijarro et al. [343] employ a game theory to propose a business model for VNOs where the network resources are outsourced to an InP and supplied to the VNOs through network slicing. Despite these efforts, extensive research is needed for developing new business models for network slices based on novel pricing and auction mechanisms that consider joint resource and revenue optimization in 5G networks. Moreover, comprehensive research is also needed to investigate fairness problem during resources allocation to 5G network slices that are requested by different MVNOs [344].

### 8.6. Mobility management in 5G network slicing

5G network slicing will face mobility management challenges caused by the increasing number of end smart-devices and different vertical industries. 5G network slices need different characteristics and requirements with regards to mobility and latency. The mobility management and handover support requirements for automated driving services is different from mobile broadband slice management. For example, over a very short period of time, high-speed trains can trigger multiple handovers for railway communications in 5G networks [345]. Fast handover with seamless mobility support is crucial for real-time services (e.g., multimedia) and has a direct influence to the end-user's QoS/QoE. However, some network slices does not need the mobility management support for 5G network slicing. For example, network slices serving industrial control do not need mobility management functions due to fixed position of devices. Recent studies has investigated the mobility management and handover mechanisms in 5G networks slicing [346–348].

Hucheng et al. [348] propose a mobility driven network slicing (MDNS) approach that takes into consideration mobility support requirements into account while customizing networks for different mobile services. Jain et al. [347] propose a Mobility Management as a Service (MMaaS) mechanism that enables the provision of globally optimized solutions for managing user mobility by allocating resources to users on-demand basis. Zhang et al. [346] introduce new mobility management schemes that can guarantee seamless handover in network-slicing-based 5G networks. Authors demonstrate that the proposed resource allocation mechanisms can allocate the available network resources between different slices in 5G systems. Moreover, authors in [349] propose an IoT-based mobility management framework that enables radio resource access to mobile roaming users across heterogeneous net-works (e.g., 5G core network and 4G evolved packet core service via the network slicing paradigm. Despite these efforts, novel approaches for mobility management have to be developed for network slicing that support service-aware QoS/QoE control in 5G systems [10]. Moreover, a seamless mobility management strategies for network slicing that can enable users to move from different SDN controllers in 5G heterogeneous systems have to be developed [350].

### 8.7. OTT-ISP collaboration for QoE-based service management in 5G network slicing

A major challenge in 5G networks is associated with the QoE provisioning to the vertical applications via network slicing. The QoE based service delivery in 5G network requires the inclusion of the QoE monitoring and QoE management concepts in the network management and orchestration paradigm [351]. In the era of the end-to-end encryption of the OTT services, the QoE monitoring and measurement requires a collaboration between the OTT and ISP/MNOs for the information exchange regarding QoE influencing factors which are in hands of OTT provider due to the fact that OTT application runs in the users terminals [47,352,353]. Regarding collaborative QoE monitoring, the approaches in [354,355] propose the installation of the passive monitoring probes with OTT applications at the UE (user terminal) and exchange of information via cloud databases with ISPs/MNOs. However, the monitoring frequency of the monitoring probe may have high impact on the overall performance in terms of the accuracy of the predicted QoE, data generated and latency in the control actions performed by the SDN controllers/MANO to ensure QoE [356]. Therefore, further studies are needed to find optimal monitoring frequency which can provide a trade off between the accuracy of the monitored QoE and latency in network operations of the end-to-end QoE-aware network slicing in collaborative network management. Furthermore, standardized interfaces to drive information-centric for the collaborative QoE-aware service management and end-to-end slicing are required. The studies in [47,353,357,358] propose architectures for the information centric collaborative QoE management in future networks. However, future studies are required to investigate scalability and effectiveness of QoE-aware collaborative service management. Moreover, standardization of the interfaces in 5G networks to drive collaborative service management is also required.

### 8.8. Summary and lesson learned

We note that, despite the recent efforts towards overcoming RAN virtualization in 5G networks, there are many challenges beyond those summarized in Table 10 such as parallelization of RAN functions, state maintenance, communication interfaces within data centers, and the impact of the RAN protocol stack. It is essential to mention that, many of these aspects are comprehensively detailed in [309,310,323]. As stated in [359], more research is needed to (1) extend the resource abstraction approach and support additional performance metrics such as latency, reliability, etc., (2) examine the performance impact on NFs dedication/sharing on different 5G network layers (3) formulate the QoS/QoE satisfaction objective when partitioning/accommodating radio resources, and (4) establish a collaboration scheme between multiple RAN run-time instances to enable large-scale control logic.

## 9. Conclusion

Both academia and industry are embracing SDN and NFV at unprecedented speed as technologies to overcome the challenge of management and orchestration of resources in 5G networks and meet different vertical's requirements. SDN and NFV promise to provide and implement new capabilities and solutions for enabling future 5G networks control and management to be adaptable, programmable and cost-effective. The concept of network slicing is the heart of 5G and will play a significant role in addressing more stringent and business-critical requirements of the vertical industries, such as real-time capabilities, latency, reliability, security and guaranteed ELAs/SLAs.

In this paper, we provide a comprehensive state-of-the-art and updated solutions related to 5G network slicing using SDN and NFV. We first present 5G service quality and business requirements followed by a description of 5G network softwarization and slicing paradigms including its concepts, history and different use cases. We then provide a tutorial of 5G network slicing technology enablers including SDN, NFV, MEC, cloud/Fog computing, network hypervisors, Virtual Machines & containers. We also comprehensively provide different industrial initiatives and projects that are pushing forward the adoption of SDN and NFV in accelerating 5G network slicing. A comparison of various 5G architectural approaches in terms of practical implementation, technology adoption and deployment strategy is given. Moreover, we provide various open source orchestrators and proof of concepts that represent an implementation from the industry. Moreover, the landscape of standardization efforts of 5G network slicing and network softwarization from both the academia and industry is highlighted.

We also present the management and orchestration of network slices in a single domain followed by a comprehensive survey of management and orchestration approaches in 5G network slicing across multiple domains while supporting multiple tenants. Also, we also provide highlights 5G network slicing management and orchestration in edge and fog networks. The last part of this paper provides future challenges and research directions related to 5G network slicing.

### Declaration of Competing Interest

The authors declare that they have no known competing financial interests or personal relationships that could have appeared to influence the work reported in this paper.

### Acknowledgment

This work was supported in part by the CONNECT Research Centre through Science Foundation Ireland, and in part by the European Regional Development Fund under Grant 13/RC/2077.

### Supplementary material

Supplementary material associated with this article can be found, in the online version, at doi:10.1016/j.comnet.2019.106984.

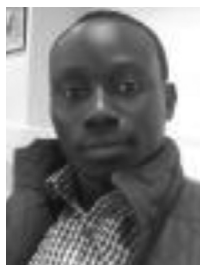

**Alcardo Alex Barakabitze** is a Research Fellow in the School of Computing, Electronics and Mathematics at the University of Plymouth, UK. He completed his PhD in QoE control and management of multimedia services in software defined and virtualized networks. He received the degree in Computer Science with Honours from the University of Dar es Salaam, Tanzania in 2010 and Master Degree of Electronics and Communication Engineering with first class from Chongqing University, PR China, in May 2015. Barakabitze is recognised as the 2015 outstanding International Graduate Student of Chongqing University, China due to his excellent performance. He was a visiting researcher in the Department of Electrical and Electronics Engineering, University of Cagliari, Italy and the ITU-T-Standardization Department in 2016 and 2017 respectively. He has numerous publication in International peer-reviewed conferences and journals. Barakabitze has served as session chair of Future Internet and NGN Architectures during the IEEE Communication Conference in Kansas City, USA. He was also the Keynote and Panel Chairs at the International Young Researcher Summit on Quality of Experience in Emerging Multimedia Services (QEEMS 2017), that was held from May 29–30, 2017 in Erfurt, Germany. Mr.Barakabitze is a Reviewer for various journals and serves on technical program committees of leading conferences focusing on his research areas. His research interests are 5G, Quality of Experience (QoE), network management, video streaming services, SDN and NFV

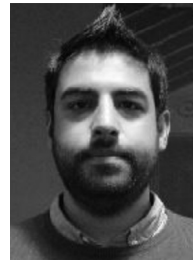

**Arslan Ahmad** is a Senior R&D Engineer at the IS-Wireless, Poland. He received his Ph.D. degree in Electronics and Computer Engineering from the Department of Electrical and Electronics Engineering, University of Cagliari, Italy in 2019. He received his engineering degree in Aviation Electronics (Avionics Engineering) in 2011 from PAF College of Aeronautical Engineering, National University of Sciences and Technology, Pakistan. In 2014, he received Master's degree in Computational Sciences and Engineering from Research Center for Modeling and Simulation, National University of Sciences and Technology, Pakistan. He has worked as Marie Curie Fellow (2015–2018) in MSCA ITN QoE-Net. He has served as a Postdoctoral researcher at University of Cagliari from 2018 to 2019. He has numerous publication in International peer-reviewed conferences and journals, two of which has been awarded Best paper award at IFIP/IEEE IM 2017 and IEEE Multimedia Technical Committee. He has been the reviewer of IEEE Transaction on Multimedia, Springer Multimedia Tools and Application, and Elsevier Image Communication. He has been Web Chair of the 10th International Conference on Quality of Multimedia Experience (QoMEX 2018) and Technical Program Committee member of IEEE ICC and IEEE WCNC. His research interests are 5G network, SDN, Quality of Experience in multimedia communication and QoE management in future Internet.

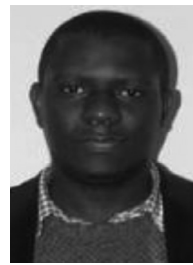

**Rashid Mijumbi** is currently a Software Systems Reliability Engineer with Nokia Bell Labs, Dublin, Ireland. He received the B.Sc. degree in electrical engineering from Makerere University, Kampala, Uganda, in 2009, and the Ph.D. degree in telecommunications engineering from the Universitat Politecnica de Catalunya (UPC), Barcelona, Spain. He was a Post-Doctoral Researcher with the UPC and the Telecommunications Software and Systems Group, Waterford, Ireland, where he participated in several Spanish national, European, and Irish National Research Projects. His current research focus is on all aspects involving future Internet, 5G, NFV, and SDN. He was a recipient of the 2016 IEEE Transactions Outstanding Reviewer Award recognizing outstanding contributions to the IEEE Transactions on Network and Service Management. He is a Reviewer for various journals and serves on technical program committees of leading conferences focusing on his research areas.

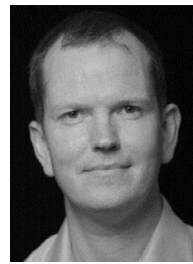

**Andrew Hines** is an Assistant Professor with the School of Computer Science, University College Dublin, Ireland. He leads the QxLab research group and is an investigator in the Science Foundation Ireland CONNECT and INSIGHT research centres, that are leaders in future networks and machine learning and data analytics respectively. He was awarded the US-Ireland Research Innovation award in 2018 by the Royal Irish Academy and American Chamber of Commerce for leading collaboration with Google since 2012 that has led to two patents, 10 industry collaborative research papers and three technology licences. The technology developed is currently used at Google for QoE modelling and testing QoE for the 30,000 most popular videos per day in YouTube. Dr Hines is a senior member of the IEEE and a leading expert in Quality of Experience for media technology. Dr Hines represented Ireland on the management committee of the European COST action on Quality of Experience, Qualinet. His primary research interests are in video signal processing, 5G, SDN and network management and machine learning for data driven quality of experience prediction across a variety of domains.